\documentclass[preprint,nofootinbib]{revtex4-1}

\usepackage[usenames,dvipsnames]{color}
\usepackage{amsmath,array,amssymb}
\usepackage{graphicx}
\usepackage{hyperref}
\usepackage{textcomp}
\usepackage{braket}
\usepackage{multirow}
\usepackage{indentfirst}
\usepackage[FIGTOPCAP,raggedright,nooneline,bf,footnotesize]{subfigure}
\usepackage{paralist}
\usepackage{overpic}
\usepackage{soul} 
\usepackage{xspace} 

\newcommand{\dd}{\ensuremath{\partial}}
\newcommand{\dt}{\ensuremath{\mathrm{d}}}
\newcommand{\gammaeit}{\ensuremath{\Gamma_\mathrm{EIT}}}
\newcommand{\gammar}{\ensuremath{\Gamma_\mathrm{R}}}

\newcommand{\tr}{\mathrm{Tr}}

\newcommand{\be}{\begin{equation}}
\newcommand{\ee}{\end{equation}}
\usepackage{bbold}
\usepackage{ulem}
\newcommand\id{\mathbb{1}}

\begin{document}

\title{Quantum optical memory protocols in atomic ensembles}

\author{Thierry Chaneli\`ere}
\affiliation{Laboratoire Aim\'e Cotton, CNRS, Univ. Paris-Sud, ENS Paris-Saclay, Universit\'e Paris-Saclay,  91405 Orsay, France}

\author{Gabriel H\'etet}
\affiliation{Laboratoire  Pierre  Aigrain,  Ecole  normale  sup\'erieure, PSL  Research  University,  CNRS,  Universit\'e  Pierre  et  Marie  Curie, Sorbonne  Universit\'es,  Universit\'e  Paris  Diderot,  Sorbonne  Paris-Cit\'e, 24  rue  Lhomond,  75231  Paris  Cedex  05,  France}

\author{Nicolas Sangouard}
\affiliation{Quantum Optics Theory Group, Department of Physics, University of Basel, CH-4056 Basel, Switzerland}


\begin{abstract}
We review a series of quantum memory protocols designed to store the {quantum} information carried by {light} into atomic ensembles. {In particular, we show how a simple semiclassical formalism allows to gain insight into various memory protocols and to highlight strong analogies between them. These analogies naturally lead to a classification of light storage protocols in{to} two categories, {namely} photon echo and {\it slow-light} memories.}  {We focus on the storage and retrieval dynamics as a key step to map the optical information into the atomic excitation.} We finally {review} various criteria adapted for both continuous variables and photon-counting measurement techniques to {certify} the quantum nature of {these} memory {protocols}.
\end{abstract}


\maketitle

\section{Introduction}

{The potential of quantum information sciences for applied physics is currently highlighted by coordinated and voluntarist policies.} In a global scheme of {probabilistic} quantum information processing, {quantum} memory is a key element {to synchronize independent events  \cite{bussieres2013prospective}}. {Memory for light} can be more generally considered as an interface between light (optical or radiofrequency) and a material medium \cite{hammerer2010quantum} where the quantum information is mapped from one form (optical for example) to the other (atomic excitation) and {\it vice versa}. {In this chapter, we review quantum protocols for light storage}. The objective is not to make a comparative and exhaustive review of the different systems {or applications} of interest. Analysis {along these lines} can be found in many review articles \cite{bussieres2013prospective, hammerer2010quantum, lvovsky2009optical, afzelius2010photon, review_Simon2010, review_Heshami_2016, review_Ma_2017} perfectly reflecting the state of the art. Instead, we focus on pioneering protocols {in atomic ensembles} that we analyze with the same formalism to extract the common {features} and differences.

{First,} we consider two representative classes of storage protocols, the photon echo in section \ref{sec:2PE} and the {\it slow-light} memories in section \ref{sec:SL}. In both cases, we first derive a minimalist semi-classical Schr\"odinger-Maxwell model to describe the propagation of a weak signal {in an atomic ensemble}. Two-level atoms are sufficient to characterize the photon echo protocols among which the standard two-pulse photon echo is the historical example (section \ref{sec:2PE}). On the contrary, as in the widely studied {\it stopped-light} by means of electromagnetically induced transparency (EIT), the minimal atomic structure consists of three levels (section \ref{sec:SL}). {In both cases}, however, the semi-classical Schr\"odinger-Maxwell formalism is sufficient to describe the optical storage dynamics and evaluate the theoretical efficiencies.

To fully replace our analysis in the context of the quantum storage, we finally derive a variety of criteria in section \ref{sec:certification} to {certify} the quantum nature of optical memories. Our approach is pragmatic in this section as we do not develop a fully quantized propagation model mirroring our semi-classical analysis in \ref{sec:2PE} and \ref{sec:SL}. Instead, we use an atomic chain quantum toy model to characterize the {noise of various storage protocols}. Criteria depending on experimentally accessible parameters are {reviewed} for both continuous and discrete variables.

\section{Photon echo memories}\label{sec:2PE}

The photon echo technique as the optical {\it alter ego} of the spin echo has been considered early as a spectroscopic tool \cite{Kopvillem,Hartmann64, Hartmann66, Hartmann68}. Its extensive description can be found in many textbooks as an example of a coherent transient light-atoms interaction \cite{allen2012optical}. Due to its coherent nature and many experimental realizations over the last decades, the photon echo has been reconsidered in the context of quantum storage \cite{afzelius2010photon}. In this section, we will first establish the formalism describing the propagation and the retrieval of week signals in a two-level inhomogeneous atomic medium. We then describe and evaluate the efficiency of the standard two-pulse photon echo from the point of view of a storage protocol. The latter is not immune to noise but has stimulated the design of noise free alternatives, namely the Controlled reversible inhomogeneous broadening and the Revival of silenced echo that we will describe using the same formalism.

The signal propagation and photon echo retrieval can be modeled by the Schr\"odinger-Maxwell equations in one dimension (along $z$) with an inhomogeneously broadened two-level atomic ensemble that we will first illustrate. 

\subsection{Two-level atoms Schr\"odinger-Maxwell model}
On one side, the atomic evolution under the field excitation is given by the Schr\"odinger equation and on the other side, the field propagation is described by the Maxwell equation that we successively remind.

\subsubsection{Schr\"odinger equation for two-level atoms}

\begin{figure}
\centering
\fbox{\includegraphics[width=.7\linewidth]{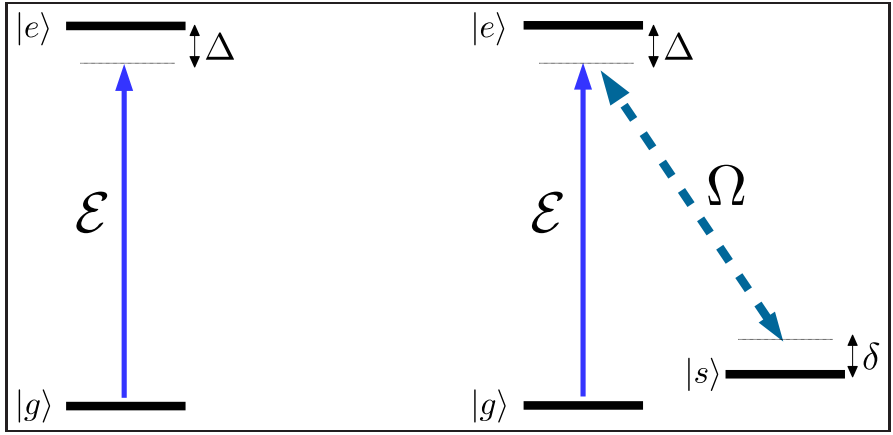}}
\caption{Two-level (left) and three-level atoms (right) used to describe the photon echo (section \ref{sec:2PE}) and the {\it slow-light} memories (section \ref{sec:SL}).The signal $\mathcal{E}$ is applied on the $|g\rangle$ and  $|e\rangle$ transition. For the three-level atoms, a control field $\Omega$ is applied on the $|s\rangle$ and  $|e\rangle$ transition.}
\label{fig:2level_3level} 
\end{figure}

For two-level atoms, labeled $|g\rangle$ and  $|e\rangle$ for the ground and excited states (see fig.\ref{fig:2level_3level}, left), the rotating-wave probability amplitudes $C_g$ and $C_e$ respectively are governed by the time-dependent Schr\"odinger equation \cite[eq. (8.8)]{shore2011manipulating}:
\begin{align}
i \partial_t \left[
\begin{array}{c}
C_g \\
C_e \\
\end{array}\right]
=
\left[
\begin{array}{ccc}
0 &\displaystyle \frac{\mathcal{E}^*}{2} \\
\displaystyle\frac{\mathcal{E}}{2} & -\Delta \\
\end{array}\right]
\left[
\begin{array}{c}
C_g \\
C_e \\
\end{array}\right]
\label{bloch2}\end{align}
where $\mathcal{E}(z,t)$ is  the complex envelope of the input signal  expressed in units of Rabi frequency. $\Delta$ is the laser detuning.

The atomic variables $C_g$ and $C_e$ depend on $z$ and $t$ for a  given detuning $\Delta$. The detunings can be made time-dependent 
\cite{loy1974observation,vitanov2001laser}, position-dependent or both \cite{hetet2008electro} but this is not the case here.

Decay terms can be added {by-hand} by introducing a complex detuning $\Delta \rightarrow \Delta - i \Gamma$ where $\Gamma$ is the decay rate of the excited state $|e\rangle$ \footnote{We do not distinguish the decay terms for the population and the coherence. This is an intrinsic limitation of the Schr\"odinger model as opposed to the density matrix formalism (optical Bloch equations).}.

\subsubsection{Maxwell propagation equation}

The propagation of the signal $\mathcal{E}(z,t)$ is described by the Maxwell equation that can be simplified in the slowly varying envelope approximation \cite[eq. (21.15)]{shore2011manipulating}. This reads for an homogeneous ensemble whose linewidth is given by the decay term $\Gamma$:
\begin{equation}
\partial_z\mathcal{E}(z,t)+ \frac{1}{c}\partial_t\mathcal{E}(z,t)=
-\displaystyle {i \alpha} \Gamma C_g^* C_e \label{MB_M_hom}
\end{equation}

The term $C_g C_e^* $ is the atomic coherence on the $|g\rangle \rightarrow |e\rangle$ transition directly proportional to the atomic polarization. The light coupling constant is included in the absorption coefficient $\alpha$ (inverse of a length unit), thus the right hand side represents the macroscopic atomic polarization.

The Maxwell equation can be generalized to an inhomogeneously broaden ensemble \cite{allen2012optical}:

\begin{equation}
\partial_z\mathcal{E}(z,t)+ \frac{1}{c}\partial_t\mathcal{E}(z,t)=
-\displaystyle\frac{i \alpha}{\pi}\int_\Delta g\left(\Delta\right) C_g^* C_e \mathrm{d}\Delta \label{MB_M_inhom}
\end{equation}

where $ g\left(\Delta\right) $ is the normalized inhomogeneous distribution.

Photon echo memories precisely rely on the inhomogeneous broadening as an incoming bandwidth. The set of equations (\ref{bloch2}\&\ref{MB_M_inhom})  are then relevant in that case. The resolution can be further simplified for weak $\mathcal{E}(z,t)$  signals as expected for quantum storage. This is the so-called perturbative regime. More importantly, the perturbative limit is necessary to ensure the linearity of the storage scheme and is then not only a formal simplification. The perturbative expansion should be used with precaution when photon echo protocols are considered. When strong (non-perturbative) $\pi$-pulses are used to trigger the retrieval as a coherence rephasing, they unavoidably invert the population. This interplay between rephasing and inversion is the essence of the photon echo technique. Population inversion should be avoided because spontaneous emission induces noise \cite{ruggiero}. We will nevertheless first consider the standard two-pulse photon echo scheme because this is the ancestor and an inspiring source for modified photon echo schemes adapted for quantum {storage}. 

\subsubsection{Coherent transient propagation in an inverted or non-inverted medium}

The goal of the present section is to describe the propagation of a weak signal representing both the incoming signal and the echo. For the standard two-pulse photon echo (see section \ref{2PE}), the echo is emitted in an inverted medium so we will consider both an inverted and a non-inverted medium corresponding to the ideal storage scheme (see \ref{CRIB} and \ref{ROSE}).
The propagation is coherent in the sense that the pulse duration is much shorter than the coherence time. The decay term (that could be introduced with a complex detuning $\Delta$) is fully neglected in eq.\eqref{bloch2}.

The coherent propagation is defined in the perturbative regime. This latter should be defined with precaution if the medium is inverted or not. The coherence term $\mathcal{P}=C_g^* C_e$ appearing in the propagation equation (eq.\ref{MB_M_hom} or \ref{MB_M_inhom}) is described by rewriting the Schr\"odinger equation
as 
\begin{equation}
\partial_t \mathcal{P} =i\Delta  \mathcal{P} + \left( C_e^* C_e-C_g^* C_g \right) i \frac{\mathcal{E}}{2}
\end{equation}

The reader more familiar with the optical Bloch equations can directly recognize the evolution of the coherence term (non-diagonal element of the density matrix) where the term $ \left( C_e^* C_e-C_g^* C_g \right) $ is the population difference (diagonal element{s}). 

For a non-inverted medium, the atoms are essentially in the ground state, so in the perturbative limit  $ \left( C_e^* C_e-C_g^* C_g \right) \rightarrow -1$. The population goes as the second order in field excitation thus justifying the perturbative expansion where the coherences $\mathcal{P}$ goes as the first order. Along the same line, $ \left( C_e^* C_e-C_g^* C_g \right) \rightarrow 1$ for an inverted medium. The atomic evolution reads as

\begin{equation}
\partial_t \mathcal{P} =i\Delta  \mathcal{P} \mp i \frac{\mathcal{E}}{2}
\end{equation}
where $\mp$ indicates if the medium is non-inverted (ground state) or inverted (excited state). This can be alternatively written in an integral form as
\begin{equation}
 \mathcal{P}(z,t) =\mp  \frac{i}{2} \int_{-\infty}^t \mathcal{E}\left(z,t^\prime\right) \exp \left(i\Delta\left(t-t^\prime\right)\right) \dt t^\prime \label{integral_form}
\end{equation}

As given by eq.\eqref{MB_M_inhom}, the propagation in the inhomogeneous medium is described by 

\begin{equation}
\partial_z\mathcal{E}(z,t)+ \frac{1}{c}\partial_t\mathcal{E}(z,t)=
-\displaystyle\frac{i \alpha}{\pi}\int_\Delta g\left(\Delta\right)  \mathcal{P}_\Delta(z,t) \dt \Delta \label{MB_M_inhom_P}
\end{equation}
We remind by an index $ \mathcal{P}_\Delta$ that the coherence term depends on the detuning $\Delta$ as a parameter.

To avoid the signal temporal distortion, the incoming pulse bandwidth should be narrower than the inhomogeneous broadening given by the distribution $g\left(\Delta\right) $ so we can safely assume $g\left(\Delta\right) \rightarrow 1$. The double integral term $\displaystyle \int_\Delta  \mathcal{P}_\Delta \dt \Delta$ from eq.\eqref{integral_form} can be simplified by writing $\displaystyle \int_\Delta  \exp \left(i\Delta\left(t-t^\prime\right)\right) \dt \Delta \rightarrow 2 \pi \delta_{t^\prime=t} $ as a representation of the Dirac peak $\delta_0$

\begin{equation}
\partial_z\mathcal{E}(z,t)+ \frac{1}{c}\partial_t\mathcal{E}(z,t)=
\mp\displaystyle\frac{\alpha}{2} \mathcal{E}(z,t)\label{bouguer0}
\end{equation}

Eq.\eqref{bouguer0}  is the absorption law or gain if the medium is inverted. The absorption law was at first discovered by Bouguer \cite{bouguer1760traite}, today known as the Bouguer-Beer-Lambert law. The description can be even more simplified by noting that the pulse length is usually much longer the medium spatial extension. The term $\displaystyle \frac{1}{c}\partial_t$ can be dropped leading to the canonical version of the Bouguer-Beer-Lambert law \cite{allen2012optical}.

\begin{equation}
\partial_z\mathcal{E}(z,t)=
\mp\displaystyle\frac{\alpha}{2} \mathcal{E}(z,t)\label{bouguer}
\end{equation}

This form can be alternatively obtained by writing the equation in the moving frame at the speed of light. Introducing the moving frame may be a source of mistake when the backward retrieval configuration is considered (see section \ref{CRIB}). {Anyway,} the moving frame {does not need to be introduced} because the medium length $L$ is in practice much shorter than the pulse extension. In other words, the delay induced by the propagation $L/c$ is negligible with respect to the pulse duration. The term {\it propagation} is in that case arguable when the term $\displaystyle \frac{1}{c}\partial_t$ is absent. Propagation should be considered in the general sense. The absorption coefficient in eq.\eqref{bouguer} defines a propagation constant. This latter is real as opposed to a propagation delay which would appear as a complex (purely imaginary) constant.

The Bouguer-Beer-Lambert law can be obtained equivalently with an homogeneous medium including the coherence decay term. This is not the case here. We insist: there is no decay and the evolution is fully coherent. To illustrate this fundamental aspect of the coherent propagation, we can show that the field excitation is actually recorded into the medium. On the contrary, with a decoherence term, the field excitation would be lost in the environment. The complete field excitation to coherence mapping is a key ingredient of the photon echo memory scheme.

\subsubsection{Field excitation to coherence mapping}\label{mapping}

In the coherent propagation regime, the evolution of the atomic and optical variables is fully coherent. Let {us} restrict the discussion to the case of interest, namely the photon echo scheme of an initially non-inverted (ground state) medium. The field is absorbed following the Bouguer-Beer-Lambert law (eq.\ref{bouguer}). This disappearance of the field is not due to the atomic dissipative decay but to the inhomogeneous dephasing. For example, in an homogeneous sample, the absorption of the laser beam can be due to spontaneous emission: the beam is depleted because the photons are scattered in other modes. In an inhomogeneous sample, the beam depletion is due to dephasing and not dissipation. In other words, the forward scattered dipole emissions destructively interfere. Since the evolution is coherent, the field should be fully mapped into the atomic excitation. In that case, the expression \eqref{integral_form} can be reconsidered by noting that after the absorption process, the integral boundary can be pushed to $+\infty$ as

\begin{align}
 \mathcal{P}_\Delta(z,t)& =-  \frac{i}{2} \exp \left(i \Delta t \right) \int_{-\infty}^{+\infty} \mathcal{E}\left(z,t^\prime\right) \exp \left(- i\Delta t^\prime \right) \dt t^\prime \\ &=-  \frac{i}{2} \exp \left(i \Delta t \right)  \tilde{\mathcal{E}}(z,\Delta) \label{mapping}
\end{align}
where $ \tilde{\mathcal{E}}(z,\omega)$  is the Fourier transform of the incoming pulse $\displaystyle \mathcal{E}\left(z,t^\prime\right)$ \footnote{ We define the Fourier transform pairs as \begin{align}
\tilde{f}(\omega)&=\int_t f(t) \exp( -i\omega t)\mathrm{d} t\\
{f}(t)&=\frac{1}{2\pi}\int_\omega \tilde{f}(\omega) \exp(i\omega t)\mathrm{d} \omega
\end{align}
}. This expression tells that the incoming spectrum is entirely mapped into the atomic excitation. More precisely, each class $\mathcal{P}_\Delta$ in the atomic distribution actually records the corresponding part in the incoming spectrum $\tilde{\mathcal{E}}(z,\Delta)$. The term $\exp \left(i \Delta t \right)$ simply reminds us that the coherence freely oscillates after the field excitation. An exponential decay term could be added by-hand by giving an imaginary part to the detuning $\Delta$.

This mapping stage when the field is recorded into the atomic coherences of an inhomogeneous medium is the initial step of the different photon echo memory schemes. Various techniques have been developed to retrieve the signal after the initial absorption stage. The inhomogeneous dephasing is the essence of the field to coherence mapping since the field spectrum is recorded in the inhomogeneous distribution. The retrieval is in that sense always associated to a rephasing or compensation of the inhomogeneous dephasing. This justifies the term photon echo used to classify this family of protocols. We will start by describing the standard two-pulse photon echo (2PE). Despite a clear limitation for quantum storage, this is an enlightening historical example. Its descendants as the so-called Controlled reversible inhomogeneous broadening (CRIB) and Revival of silenced echo (ROSE) have been precisely designed to avoid the deleterious effect of the $\pi$-pulse rephasing used in the 2PE sequence.

\subsection{Standard two-pulse photon echo}\label{2PE}
 
\begin{figure}
\centering
\fbox{\includegraphics[width=.75\linewidth]{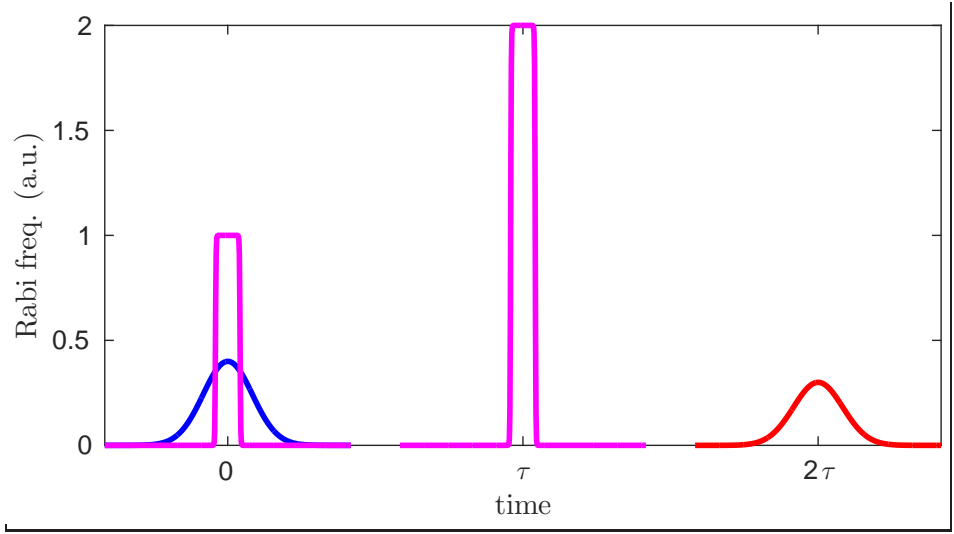}}
\caption{Standard two-pulse photon echo sequence. Inherited from magnetic resonance, $\pi/2-\pi$ sequence (delayed by $\tau$, in magenta) produces an echo  at $2\tau$ (in red). When considered for optical storage, the first pulse is weak (in blue) and longer than the  rephasing $\pi$-pulse.}
\label{fig:2PE}
\end{figure}

Inherited from the magnetic resonance technique \cite{hahn}, the coherence rephasing {and the subsequent field reemission} is triggered by applying a strong $\pi$-pulse (fig.\ref{fig:2PE}). The possibility to use the 2PE for pulse storage has been mentioned early in the context of optical processing \cite{Carlson:83}. The retrieval efficiency can indeed be remarkably high \cite{moiseev1987some,azadeh, SjaardaCornish:00}.
 
This particularity has attracted a renewed curiosity in the context of quantum information \cite{rostovtsev2002photon, moiseev_echo_03}.

\subsubsection{Retrieval efficiency}

The retrieval efficiency can be derived analytically from the Schr\"odinger-Maxwell model. Following the sequence in fig.\ref{fig:2PE}, the signal absorption is first described by the Bouguer-Beer-Lambert law (eq.\ref{bouguer}). The initial stage is followed by a free evolution during a delay $\tau$. The $\pi$-pulse will trigger a retrieval. The action of a strong pulse on the atomic variables is described by the propagator

\begin{align}
\left[
\begin{array}{c}
C_g\left(\tau^+\right) \\
C_e\left(\tau^+\right) \\
\end{array}\right]
=
\left[
\begin{array}{ccc}
\displaystyle \cos(\theta/2) & \displaystyle -i \sin(\theta/2) \\
\displaystyle -i \sin(\theta/2) &\displaystyle  \cos(\theta/2) \\
\end{array}\right]
\left[
\begin{array}{c}
C_g \left(\tau^-\right) \\
C_e \left(\tau^-\right) \\
\end{array}\right]
\end{align}
which links the atomic variables just before ($\tau^-$) and after  ($\tau^+$) a general $\theta$-area pulse. This solution of the canonical Rabi problem is only valid for a {very short} pulse (hard pulse). More precisely, in the atomic evolution eq.\eqref{bloch2}, the Rabi frequency {must be} much larger than the detuning. In the 2PE scheme, this means that the atoms excited by the signal (first pulse) are uniformly (spectrally) covered by the strong rephasing pulse. This translates in the time domain as a condition on the relative pulse durations: the $\pi$-pulse {must} be much shorter than signal. This aspect appears as an initial condition for the pulse durations but is also intimately related to the transient coherent propagation of strong pulses among which $\pi$-pulses are a particular case. This will be discussed in more details in the appendix \ref{strong_pulse}. Assuming the ideal situation of the uniform $\theta=\pi$ pulse area, the propagator takes the simple form
$$
\left[
\begin{array}{ccc}
 0 & -i \\
-i & 0 \\
\end{array}\right]$$
fully defining the effect of the $\pi$-pulse on the stored coherence

\begin{equation}
\mathcal{P}_\Delta\left(\tau^+\right)=-\mathcal{P}_\Delta^*\left(\tau^-\right) =-  \frac{i}{2} \exp \left(-i \Delta \tau \right)  \tilde{\mathcal{E}}^*(z,\Delta)\label{P_tauplus}
\end{equation}

{T}he free evolution resumes by adding the  inhomogeneous phase $\Delta \left( t- \tau\right)$

\begin{equation}
\mathcal{P}_\Delta\left(t>\tau\right)=\mathcal{P}_\Delta\left(\tau^+\right) \exp \left(i \Delta \left( t- \tau\right) \right) =-  \frac{i}{2} \exp \left(i \Delta \left( t- 2\tau\right) \right)  \tilde{\mathcal{E}}^*(z,\Delta)\label{P_echo}
\end{equation}
In the expression \eqref{P_echo}, we see that the inhomogeneous phase $\Delta \left( t- 2\tau\right)$ is zero at the instant $t=2\tau$ of the retrieval thus justifying the term rephasing.

The propagation of the retrieved echo $\mathcal{E}^R$ follows eq.\eqref{MB_M_inhom_P}. The source term on the right-hand side has now two contributions.  The first one gives the Bouguer-Beer-Lambert law (eq \ref{bouguer}) for the echo field $\mathcal{E}^R$ itself. A critical aspect of the 2PE is the population inversion induced by the $\pi$-pulse. The intuition can be confirmed by calculating from the propagator $ \left( C_e^*\left(\tau^+\right) C_e\left(\tau^+\right)-C_g^*\left(\tau^+\right) C_g\left(\tau^+\right) \right)$ to the first order by noting that $ C_g\left(\tau^-\right) \simeq 1$. The echo field $\mathcal{E}^R$ exhibits gain. The second one comes from the coherence initially excited by the signal freely oscillating after the $\pi$-pulse rephasing. In other words, the coherences at the instant of retrieval are the sum of the free running term due to the signal excitation from eq.\eqref{P_echo} and the contribution from the echo field itself.

\begin{equation}
\partial_z\mathcal{E}^R(z,t)=
+\displaystyle\frac{\alpha}{2} \mathcal{E}^R(z,t) -\displaystyle\frac{i \alpha}{\pi}\int_\Delta g\left(\Delta\right)  \mathcal{P}_\Delta(z,t>\tau) \dt \Delta
\end{equation}
The integral source term representing the build-up of the macroscopic polarisation at the instant of retrieval is directly related to the signal field excitation $\mathcal{E}$ which appears as the inverse Fourier transform of $\tilde{\mathcal{E}}^*(z,\Delta)$ from eq.\eqref{P_echo}, {that is} 

\begin{equation}
\partial_z\mathcal{E}^R(z,t)=
+\displaystyle\frac{\alpha}{2} \mathcal{E}^R(z,t) - \alpha \mathcal{E}^*(z,2\tau-t) \label{eq_echo}
\end{equation}

{Eq.\eqref{eq_echo} is simple but rich because it can be modified by-hand to describe the descendants of the 2PE protocol that are suitable for quantum storage as we will see in sections \ref{CRIB} and \ref{Rose}. Note that it can be adapted to account for rephasing pulse areas $\theta$ that are not $\pi.$ They lead to imperfect rephasing and incomplete medium inversion thus modifying the terms in eq.\eqref{eq_echo} \cite{ruggiero}. Very general expressions for the efficiency as a function of $\theta$ can be analytically derived \cite{moiseev1987some}.}
{Knowing that the incoming signal follows the Bouguer-Beer-Lambert law (eq.\ref{bouguer}) of absorption $\displaystyle \mathcal{E}(z,t)=\mathcal{E}(0,t) \exp\left(-\alpha z/2\right),$ the efficiency of the 2PE can be obtained as a function of optical depth $d=\alpha L$ from the ratio between the output and input intensities
\begin{equation}
\eta=\frac{|\mathcal{E}^R(L,t)|^2}{|\mathcal{E}(0,2\tau-t)|^2}\label{efficiency}
\end{equation}
For a $\pi$-rephasing pulse, we find 
\begin{equation}\label{etaPi}
\eta\left(d\right)=\left[\exp\left(d/2\right)-\exp\left(-d/2\right)\right]^2 =4~{\rm sinh}^2\left(d/2\right)
\end{equation}
}
{At large optical depth $d$, the efficiency scales as $\exp\left(d\right)$ resulting in an exponential amplification of the input field. This amplification prevents the 2PE to be used as a quantum storage protocol. The simplest but convincing argument uses the no-cloning theorem \cite{nocloning}. Alternatively, we can apply various criteria to certify the quantum nature of the memory on the echo and show that none of these criteria witnesses its non-classical feature, as wee will see section \ref{sec:certification}.}

\begin{figure}
\centering
\fbox{\includegraphics[width=.85\linewidth]{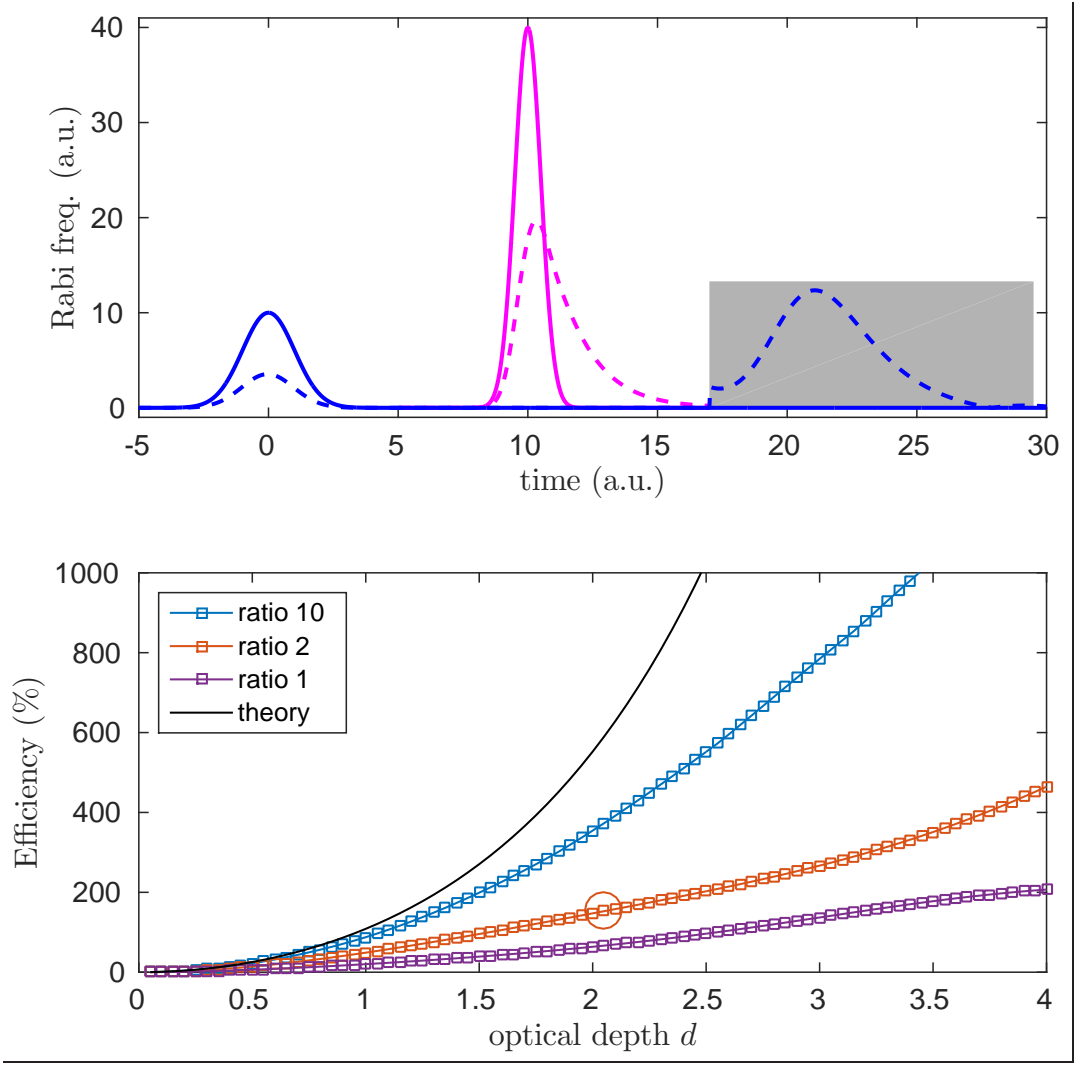}}
\caption{Top: Numerical simulation of a 2PE sequence with a weak incoming signal (area $\pi/20$). The signal and the echo fields are in blue and have been magnified by a factor 10 (shaded area). The incoming pulses are in solid lines. The outgoing pulses after propagation though $d = 2$ are in dashed line. The $\pi$-pulse is two times shorter than the signal. Bottom: Storage efficiencies (see text for the definition) as a function of the optical depth $d$ . The black line is the analytical solution eq.\eqref{etaPi}. Three simulations have been performed depending on the relative duration of the $\pi$-pulse with respect to the signal: when the $\pi$-pulse has the same duration than the signal (ratio 1), when it is 2 times (ratio 2) and 10 times shorter (ratio 10). The circle corresponds to the 2PE sequence on top.}
\label{fig:2PE_simul}
\end{figure}

In fig.\ref{fig:2PE_simul} (bottom), we have represented this efficiency scaling (eq.\ref{etaPi}) that we compare with a numerical simulation of a 2PE sequence solving the Schr\"odinger-Maxwell model. For a given inhomogeneous detuning $\Delta$, we calculate the atomic evolution eq.\eqref{bloch2}  by using a fourth-order Runge-Kutta method. After summing over the inhomogeneous broadening, the output pulse is obtained by integrating eq.\eqref{MB_M_inhom} along $z$ using the Euler method.

In the numerical simulation, there is no assumption on the $\pi$-pulse duration with respect to the signal bandwidth (as needed to derive the analytical formula eq.\ref{etaPi}). The excitation pulses are assumed Gaussian as shown for the incoming and the outgoing pulses of a 2PE sequence after propagation though an optical depth $d= 2$ (fig.\ref{fig:2PE_simul}, top). We consider different durations for the $\pi$-pulse (of constant area) and a fixed signal duration.

From the numerical simulation, the efficiency is evaluated by integrating under the intensity curves of the echo (shades area). This latter reaches 152\% for the sequence of fig.\ref{fig:2PE_simul} (top), larger than 100\% as expected for an inverted medium. Still, this is much smaller than the 552\% {efficiency} expected from eq.\eqref{etaPi} with $d= 2$. This discrepancy is essentially explained by the $\pi$-pulse distortion through propagation (magenta dashed line  in fig.\ref{fig:2PE_simul},top) than can be observed numerically. The $\pi$-pulse should stay shorter than the signal to properly ensure the coherence rephasing. This is obviously not the case because the pulse is distorted as we briefly analyze in appendix \ref{strong_pulse} with the energy and area conservation laws.

As a summary, we have evaluated numerically the efficiencies when the $\pi$-pulse has the same duration than the signal (ratio 1), when it is 2 times and 10 times shorter (ratio 2 and 10 respectively). We see in  fig.\ref{fig:2PE_simul} (bottom) than the efficiencies deviates significantly from the prediction eq.\eqref{etaPi}. There is less discrepancy when the  $\pi$-pulse is 10 times shorter than the signal (ratio 10), especially at low optical depth. Still, for larger $d$, the distortions are sufficiently important to reduce the efficiency significantly.

Despite a clear deviation from the analytical scaling (eq.\ref{etaPi}), the echo amplification is important (efficiency $>$ 100\%). This latter comes from the inversion of the medium. As a consequence, the amplified spontaneous emission mixes up with the retrieved signal then inducing noise. It should be noted that the signal to noise ratio only depends on the optical depth \cite{ruggiero,RASE,Sekatski}. This may be surprising at first sight because the coherent emission of the echo and the spontaneous emission seems to have completely different collection patterns offering a significant margin to the experimentalist to filter out the noise. This is not the case. The excitation volume is defined by the incoming laser focus. On the one hand, a tighter focus leads to a smaller number of inverted atoms thus reducing the number spontaneously emitted photons. On the other hand, a tight focus requires a larger collection angle of the retrieved echo. Less atoms are excited but the spontaneous emission collection angle is larger. The noise in the echo mode is unchanged. This qualitative argument which can be seen as a conservation of the optical etendue is quantitatively supported by a quantized version of the Bloch-Maxwell equations \cite{RASE,Sekatski}. This aspect will be discussed in sections \ref{CV_criterion} and \ref{counting_criterion} using a simplified quantum model.

In any case, population inversion should be avoided. This statement motivated many groups to conceive rephasing protocols by keeping the best of the 2PE but avoiding the deleterious effect of $\pi$-pulses as we will see now in sections \ref{CRIB} and \ref{Rose}.

\subsection{Controlled reversible inhomogeneous broadening}\label{CRIB}

\begin{figure}
\centering
\fbox{\includegraphics[width=.75\linewidth]{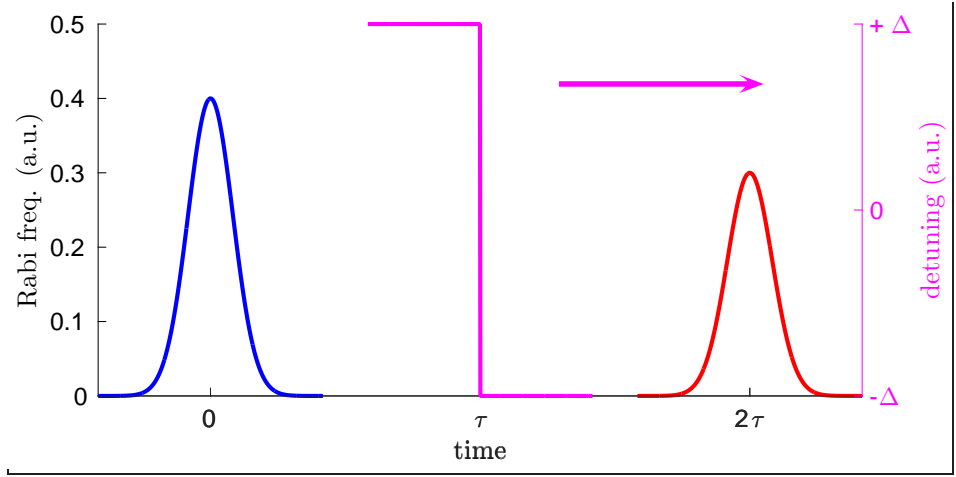}}
\caption{CRIB echo sequence. As compared to the 2PE sequence, no rephasing pulse is applied, the inhomogeneous broadening is reversed using a controllable electric field for example \cite{CRIB3} (magenta line).}
\label{fig:CRIB}
\end{figure}

The controlled reversible inhomogeneous broadening (CRIB) offers a solid alternative to the 2PE \cite{CRIB1,CRIB2,CRIB3,CRIB4, sangouard_crib}. The CRIB,,as represented	 in fig.\ref{fig:CRIB},  has been successfully implemented with large efficiencies \cite{hedges2010efficient, hosseini2011high} and low noise measurements \cite{lauritzen2010telecommunication} validating the protocol as a quantum memory in different systems, from atomic vapors to doped solids \cite{lauritzen2011approaches}.

Fundamentally, an echo is generated by rephasing the coherences corresponding to the cancellation of the inhomogeneous phase. As indicated by eq.\eqref{mapping}, the accumulated phase is $\Delta t$. Taking control of the detuning $\Delta$ is sufficient to produce an echo without a $\pi$-pulse. This is the essence of the CRIB sequence, where the detuning is actively switched from $\Delta$ for $t<\tau$ to  $-\Delta$ for $t>\tau$. We won't focus on the realization of the detuning inversion. This aspect has been covered already and we recommend the reading of the review papers \cite{lvovsky2009optical, afzelius2010photon}. We here focus on the coherence rephasing and evaluate the efficiency which can be compared to other protocols. It should be noted that the gradient echo memory scheme (GEM) \cite{hetet2008electro} is not covered by our description. We will assume that the coherences undergo the transform $+\Delta \rightarrow -\Delta$ independently of the atomic position $z$. This is not the case for the GEM where the detuning $\Delta$ goes linearly (or at least monotonically) with the position $z$. The GEM can be called the longitudinal CRIB. This specificity of the GEM makes it remarkably efficient \cite{hetet2008electro, hedges2010efficient, hosseini2011high}.

\begin{figure}
\centering
\fbox{\includegraphics[width=.75\linewidth]{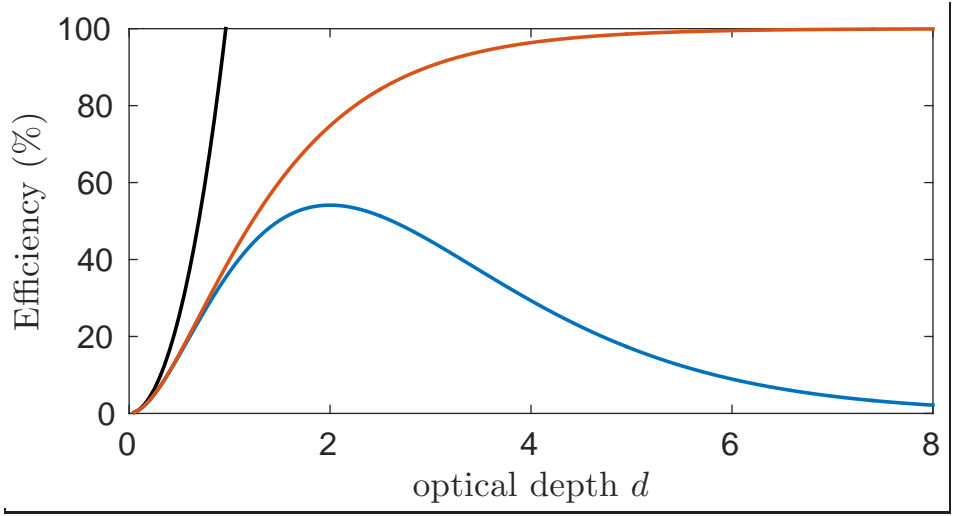}}
\caption{Comparison of the forward (eq.\ref{eta_crib}, in blue) and backward (eq.\ref{eta_crib_back}, in red) CRIB efficiency scaling. The standard 2PE efficiency is represented as a reference (eq.\ref{etaPi}, in black)}
\label{fig:compar_eff}
\end{figure}

Assuming that $\Delta \rightarrow -\Delta$ for $t>\tau$, it should be first noted that at the switching time $\tau$, the coherence term is continuous

\begin{equation}
\mathcal{P}_\Delta\left(\tau^+\right)=\mathcal{P}_\Delta\left(\tau^-\right) =-  \frac{i}{2} \exp \left(i \Delta \tau \right)  \tilde{\mathcal{E}}(z,\Delta)
\end{equation}
but will evolve with a different detuning afterward, that is
\begin{equation}
\mathcal{P}_\Delta\left(t>\tau\right)=\mathcal{P}_\Delta\left(\tau^+\right) \exp \left(-i \Delta \left( t- \tau\right) \right) =-  \frac{i}{2} \exp \left(i \Delta \left(2\tau -t\right) \right)  \tilde{\mathcal{E}}(z,\Delta)\label{P_crib}
\end{equation}
The latter gives the source term of the differential equation defining the efficiency similar to eq.\eqref{eq_echo} for the 2PE
\begin{equation}
\partial_z\mathcal{E}^R(z,t)=
-\displaystyle\frac{\alpha}{2} \mathcal{E}^R(z,t) - \alpha \mathcal{E}(z,2\tau-t) \label{eq_crib}
\end{equation}
Eqs \eqref{eq_echo} and \eqref{eq_crib} are very similar. The first term on the right hand side is now negative (proportional to $-\displaystyle\frac{\alpha}{2}$)  because the medium is not inverted in the CRIB sequence. This is a major difference. Again, the incoming signal follows the Bouguer-Beer-Lambert law of absorption $\displaystyle \mathcal{E}(z,t)=\mathcal{E}(0,t) \exp\left(-\alpha z/2\right)$ but the efficiency defined by \eqref{efficiency} is now {given} after integration by
\begin{equation}\label{eta_crib}
\eta\left(d\right)=d^2 \exp\left(-d\right)
\end{equation}
The maximum efficiency is obtained for $d=\alpha L=2$ with $\eta\left(2\right)=54\%$ \cite{sangouard_crib} (see fig.\ref{fig:compar_eff}). There is no gain so the semi-classical efficiency is always smaller than one. The efficiency is limited in the so-called forward configuration because the echo is {\it de facto} emitted in an absorbing medium. The re-absorption of the echo limits the efficiency to $54\%$. Ideal echo emission with unit efficiency can be obtained in the backward configuration. This latter is implemented by applying auxiliary pulses, typically Raman pulses modifying the phase matching condition from forward to backward echo emission. The Raman pulses increase the storage time by shelving the excitation into nuclear spin state for example. This ensures the complete reversibility by flipping the apparent temporal evolution (as shown by eq.\eqref{P_crib}) and the wave-vector \cite{reversibility}.

Despite its simplicity, eq.\eqref{eq_crib} can be adapted to describe the backward emission without working out the exact phase matching condition. We consider the following equivalent situation. The signal is first absorbed: $\displaystyle \mathcal{E}(z,t)=\mathcal{E}(0,t) \exp\left(-\alpha z/2\right)$. We now fictitiously flip the atomic medium: the incoming slice $z=0$ becomes $z=L$ and {\it vice versa}. The atomic excitation would correspond to the absorption of a  backward propagating field
$$\displaystyle \mathcal{E}(z,t)=\mathcal{E}(0,t) \exp\left(\alpha \left(z-L\right)/2\right)$$
Eq.\eqref{eq_crib} can be integrated with this new boundary condition, giving the backward efficiency of the CRIB
\begin{equation}\label{eta_crib_back}
\eta\left(d\right)=\left[1 - \exp\left(-d\right)\right]^2
\end{equation}
For a sufficiently large optical depth, the efficiency is close to unity. As a comparison, we have represented the forward (eq.\ref{eta_crib}) and backward (eq.\ref{eta_crib_back}) CRIB efficiencies in fig.\ref{fig:compar_eff}.

 The practical implementation of the CRIB requires to control dynamically the detuning by Stark or Zeeman effects. The {\it natural} inhomogeneous broadening has a static microscopic origin and cannot be used {\it as it is}. The initial optical depth has to be sacrificed to obtain an effective controllable broadening. This statement motivates the reconsideration of the 2PE which precisely exploit the bare inhomogeneous broadening offering advantages in terms of available optical depth and bandwidth.

\subsection{Revival of silenced echo}\label{ROSE} \label{Rose}

\begin{figure}
\centering
\fbox{\includegraphics[width=.85\linewidth]{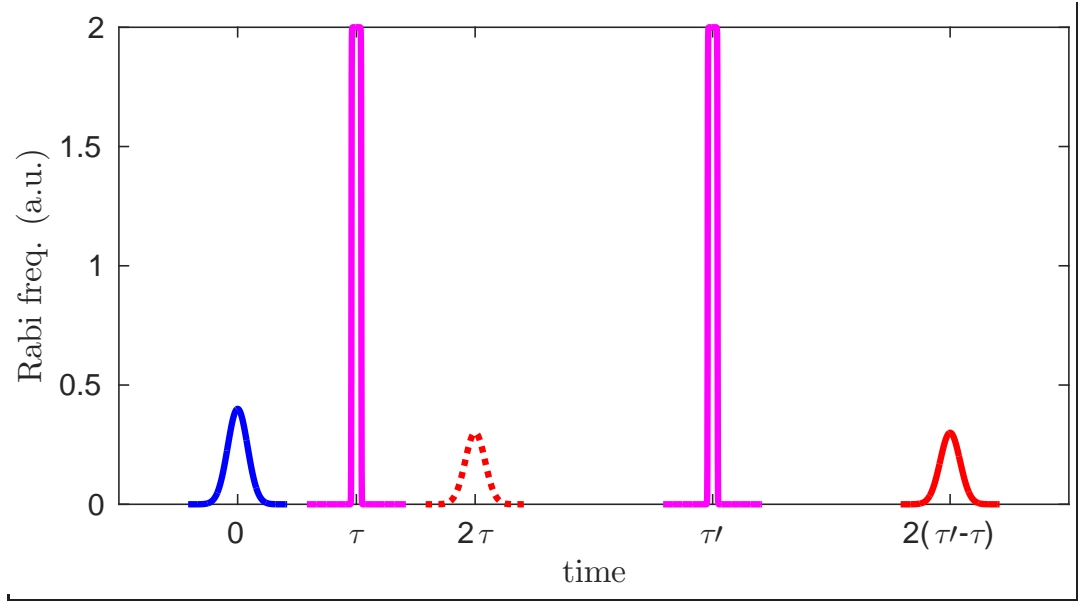}}
\caption{A schematic ROSE echo sequence that can be seen as the concatenation of two 2PE sequences (fig.\ref{fig:2PE}). The first echo at $t=\tau$ (in dashed red) should be silenced by the phase matching conditions (see \ref{phase_match}).  A second $\pi$-pulse at $t=\tau^\prime$ induces the emission of the ROSE echo at $t=2\left(\tau^\prime-\tau\right)$ (in red).}
\label{fig:ROSE}
\end{figure}

The Revival of silenced echo (ROSE) is a direct descendant of the 2PE \cite{rose}. The ROSE is essentially a concatenation of two 2PE sequences as represented in fig.\ref{fig:ROSE}. In practice, the ROSE sequence advantageously replace{s} $\pi$-pulses by complex hyperbolic secant (CHS) pulses as we will specifically discuss in \ref{strong_pulse_rose}. For the moment, we assume that the rephasing pulses are simply $\pi$-pulses. This is sufficient to evaluate the efficiency and derive the phase matching conditions.

Concatenated with a 2PE sequence, a second $\pi$-pulse (at $t=\tau^\prime$ in fig.\ref{fig:ROSE}) triggers a second rephasing of the coherences at $t=2\left(\tau^\prime-\tau\right)$. This latter leaves the medium non-inverted avoiding the deleterious effect of a single 2PE sequence. This reasoning is only valid if the first echo is not emitted. In that case, the coherent free evolution continues after the first rephasing. The first echo is said to be silent (giving the name to the protocol) because the coherence rephasing is not associated to a field emission. The phase matching conditions are indeed designed to make the first echo silent but preserve the final retrieval of the signal. Along the same line with the same motivation, McAuslan {\it et al.} proposed to use the Stark effect to silence the emission of the first echo \cite{HYPER} by cunningly applying the tools developed for the CRIB to the 2PE, namely by inducing an artificial inhomogeneous reversible broadening. The AC-Stark shift (light shift) also naturally appeared as a versatile tool to manipulate the retrieval \cite{Chaneliere:15}. We will discuss the phase matching conditions latter. Before that, we will evaluate the retrieval efficiency applying the method developed for the 2PE and CRIB.

\subsubsection{Retrieval efficiency}

Following the procedure in section \ref{2PE}, we assume that a second  $\pi$-pulse is applied at $t=\tau^\prime$. Starting from eq.\eqref{P_echo}, we can track the inhomogeneous phase at $t=\tau^\prime$ when the $\pi$-pulse is applied (similar to eq.\ref{P_tauplus}) as
\begin{equation}
\mathcal{P}_\Delta\left(\tau^{\prime+}\right)=-\mathcal{P}_\Delta^*\left(\tau^{\prime-}\right) =-  \frac{i}{2} \exp \left(-i \Delta \left( \tau^\prime -2\tau\right) \right)  \tilde{\mathcal{E}}^*(z,\Delta)
\end{equation}
freely evolving afterward as
\begin{equation}
\mathcal{P}_\Delta\left(t>\tau^\prime\right)=-  \frac{i}{2} \exp \left(i \Delta \left( t- 2\tau^\prime+2\tau\right) \right)  \tilde{\mathcal{E}}(z,\Delta)\label{P_rose}
\end{equation}
There is indeed a rephasing at $ t=2\left(\tau^\prime-\tau\right)$. The retrieval follows the common  differential equation (as eqs. \eqref{eq_echo} and \eqref{eq_crib})
\begin{equation}
\partial_z\mathcal{E}^R(z,t)=
-\displaystyle\frac{\alpha}{2} \mathcal{E}^R(z,t) - \alpha \mathcal{E}(z,t- 2\tau^\prime+2\tau) \label{eq_rose}
\end{equation}
As compared to the 2PE, the ROSE echo is not emitted in an inverted medium. One can note that the signal is not time-reversed as in the 2PE and CRIB, so the efficiency is defined as
\begin{equation}
\eta=\frac{|\mathcal{E}^R(L,t)|^2}{|\mathcal{E}(0,t- 2\tau^\prime+2\tau)|^2}
\end{equation}
The ROSE efficiency is exactly similar to CRIB due to the similarity of eqs.\eqref{eq_crib} and \eqref{eq_rose}. It is limited to 54\% in the forward direction because the medium is absorbing. {Complete reversal can be obtained in the backward direction by precisely designing the phase matching condition, the latter being a critical ingredient of the ROSE protocol.}

Even if there is no population inversion at the retrieval, the use of strong pulses for the rephasing is a potential source of noise. First of all, any imperfection of the $\pi$-pulses may leave some population in the excited state leading to a partial amplification of the signal. Secondarily, the interlacing of strong and weak pulses within the same temporal sequence is like playing with fire. This is a common feature of many quantum memory protocols for which control fields may leak in the signal mode. Many experimental techniques are combined to isolate the weak signal: different polarization, angled beams (spatial selection) and temporal separation. Encouraging demonstrations of the ROSE down to few photons per pulses have been performed by combining theses techniques \cite{bonarota_few}, thus showing the {potentials} of the protocol.

\subsubsection{Phase matching conditions}\label{phase_match}

Phase matching can be considered in a simple manner by exploiting the spectro-spatial analogy. Each atom in the inhomogeneous medium is defined by its detuning (frequency) and position (space), both contributing to the inhomogeneous phase. In that sense, the instant of emission can be seen as a spectral phase matching condition.  Following this analogy, the spatial phase matching condition can be derived from the photon echo time sequence \cite{mukamel}.

Let {us} take the 2PE as an example (fig.\ref{fig:2PE}). The 2PE echo is emitted at $t=t_1+2\tau=2t_2-t_1$ where $t_1$ is the arrival time of the signal (first pulse) and $t_2$ the $\pi$-pulse (second pulse). In fig.\ref{fig:2PE}, we have chosen $t_1=0$ and $\tau=t_2-t_1$ for simplicity . By analogy, the echo should be emitted in the direction $\overrightarrow{k}=2\overrightarrow{k_2}-\overrightarrow{k_1}$ where $\overrightarrow{k_1}$ and $\overrightarrow{k_2}$ are the wavevectors of the signal and $\pi$-pulse respectively. In that case, if $\overrightarrow{k_1}$ and $\overrightarrow{k_2}$ are not collinear ($\overrightarrow{k_1}\neq \overrightarrow{k_2}$), the phase matching cannot be fulfilled: there is no 2PE echo emission.

\begin{figure}
\centering
\fbox{\includegraphics[width=.8\linewidth]{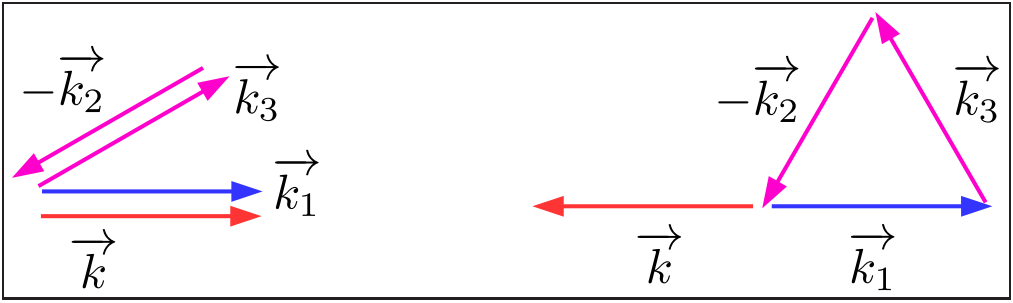}}
\caption{ROSE Phase matching conditions. Left: canonical experimental situation where the two $\pi$-pulses are on the same beam. The echo $\vec{k}$  is in the signal mode $\vec{k_1}$ (forward). The angle between the signal and the rephasing can be large for a good isolation of the echo as recently tested in the orthogonal configuration \cite{Gerasimov2017}. Right: backward retrieval of the ROSE echo $\vec{k}$. The signal $\vec{k_1}$ and two rephasing beams forms a equilateral triangle in that case: the echo is emitted backward.}
\label{fig:phase_matching}
\end{figure}

Following the same procedure, the ROSE echo is emitted at $t=t_1+2(\tau^\prime-\tau)=t_1+2(t_3-t_2)$ where $t_3$ is the arrival time of the second $\pi$-pulse (third pulse). The ROSE echo should be emitted if the $\overrightarrow{k}=\overrightarrow{k_1}+2(\overrightarrow{k_3}-\overrightarrow{k_2})$ direction ($\overrightarrow{k_3}$ is the direction of the second $\pi$-pulse). The canonical experimental situation satisfying the ROSE phase matching condition corresponds to   $\overrightarrow{k_1}\neq \overrightarrow{k_2}$ (not collinear) but keeping $\overrightarrow{k_3}=\overrightarrow{k_2}$ \cite{Dajczgewand:14, Gerasimov2017}. There is no 2PE in that case because  $\overrightarrow{k_1}\neq \overrightarrow{k_2}$ but the ROSE echo is emitted in the direction $\overrightarrow{k_1}$ of the signal as represented in fig.\ref{fig:phase_matching}.

The backward retrieval configuration is illustrated as well in fig.\ref{fig:phase_matching} (right). The efficiency can reach 100\% because the reversibility of the process is ensured spatially and temporally.

\subsubsection{Adiabatic pulses}\label{strong_pulse_rose}
Even if the protocol can be understood with $\pi$-pulses, the rephasing pulses can be advantageously replaced by complex hyperbolic secant (CHS) in practice \cite{Dajczgewand:14, Gerasimov2017}. The CHS are another heritage from the magnetic resonance techniques \cite{Garwood2001155}. As representative of the much broad class of adiabatic and composite pulses, CHS produce a robust inversion because for example the final state weakly depends on the pulse shape and amplitude. Within a spin or photon echo sequence, they must be applied by pairs because each CHS adds an inhomogeneous phase due to the frequency sweep. This latter can be interpreted as a sequential flipping of the inhomogeneous ensemble. Two identical CHS produce a perfect rephasing because the inhomogeneous phases induced by the CHS cancel each other \cite{minar_chirped, PascualWinter}.

CHS additionally offer{s} an advantage that is somehow underestimated. As we have just said, CHS must be appl{ied} by pairs. It means that the first echo in the ROSE sequence is also silenced because it would follow the first CHS, as opposed to the second echo which follows a pair of CHS. How much the first echo is silenced depends on the parameters of the CHS, namely the Rabi frequency and the frequency sweep. This degree of freedom should not be neglected when the phase matching conditions cannot be modified as in the cavity case in the optical or RF domain \cite{Grezes}.

To conclude about the ROSE and because of its relationship with the 2PE, it is important to question the strong pulse propagation that we pointed out as an important efficiency limitation of the 2PE (with $\pi$-pulses) by analyzing fig.\ref{fig:2PE_simul} (see appendix \ref{strong_pulse} for a more detailled discussion). In that sense as well, the CHS are superior to $\pi$-pulses. CHS are indeed very robust to propagation in absorbing media so their preserve their amplitude and frequency sweep \cite{Warren, PhysRevLett.82.3984}. CHS are not constrained by the McCall and Hahn Area Theorem (eq.\ref{area}). The latter isn't valid for frequency swept pulses \cite{Eberly:98}. This robustness to propagation can be explained qualitatively by considering the energy conservation \cite{rose}.

The different advantages of the CHS as compared to $\pi$-pulses have been studied accurately using numerical simulations in \cite{Demeter}, confirming both their versatility and robustness.

\subsection{Summary and perspectives}

We have described the variations from the well-known photon echo technique adapted for quantum storage. We haven't discussed in details the gradient echo memory scheme (GEM) \cite{hetet2008electro} (sometimes called longitudinal CRIB) which can be seen as an evolution of the CRIB protocol. The GEM is remarkable for its efficiency \cite{hetet2008electro, hedges2010efficient, hosseini2011high} allowing demonstrations in the quantum regime of operation \cite{hosseini2011unconditional}. The scheme has been enriched {by} processing functions as {a} pulse sequencer \cite{hosseini2009coherent, hosseini_jphysb}. More importantly, the GEM has been considered for RF storage in an ensemble of spins thus covering different physical realities and frequency ranges \cite{wu2010storage, zhang2015magnon}. As previously mentioned, the GEM is not covered by our formalism because the scheme couples the detuning and the position $z$. An analytical treatment is possible but is beyond the scope of our paper \cite{LongdellAnalytic}.

The specialist reader may be surprised because we did not discussed the atomic frequency comb (AFC) protocol \cite{afc} despite an undeniable series of success. The early demonstration of {weak classical field and single photon} storage \cite{usmani2010mapping, saglamyurek2011broadband, clausen2011quantum, PhysRevLett.108.190505, gundogan, PhysRevLett.115.070502, Tiranov:15, maring2017photonic} has been pushed to a remarkable level of integration \cite{saglamyurek2015quantum, PhysRevLett.115.140501, Zhong1392}. The main advantage of the AFC is a high multimode capacity \cite{afc, bonarota2011highly} which has been identified as an critical feature of the deployment of quantum repeaters \cite{collins,simon2007}. Despite a clear filiation of the AFC with the photon echo technique \cite{Mitsunaga:91}, there are also fundamental differences. For the AFC, there is no direct field to coherence mapping as discussed in section \ref{mapping}. The AFC is actually based on a population grating. Without going to much into a semantic discussion, the AFC is a descendant of the three-pulse photon echo and not the two-pulse photon echo \cite{mukamel} that we analyze in this section \ref{sec:2PE}. As a consequence, the AFC can be surprisingly linked to the {\it slow-light} protocols \cite{afc_slow} that we will discuss in the next section \ref{sec:SL}


\section{Slow-light memories}\label{sec:SL}
Since the seminal work of Brillouin \cite{brillouin} and Sommerfeld \cite{sommerfeld},  {\it slow-light} is a fascinating subject whose impact has been significantly amplified by the popular science-fiction culture \cite{shaw}. The external control of the group velocity reappeared in the context of quantum information as a mean to store and retrieve optical {light while preserving its quantum features} \cite{EIT_Harris, fleischhauer2000dark, FLEISCHHAUER2000395}. The rest is a continuous success story that can only be embraced by review papers \cite{review_Ma_2017}.

We will start this section by deriving the Schr\"odinger-Maxwell equations used to describe the signal storage and retrieval. Our analysis is based on the following classification. We first consider the fast storage and retrieval scheme as introduced by Gorshkov {\it et al.} \cite{GorshkovII}. In other words, the storage is triggered by brief Raman $\pi$-pulses \cite{GorshkovII, legouet_raman}. We then consider the more established electromagnetically induced transparency (EIT) and the Raman schemes. In theses cases, the storage and retrieval are activated by a control field that is on or off. The difference between EIT and Raman is the control field detuning: on-resonance for the EIT scheme and off-resonance for the Raman. Both lead to very different responses of the atomic medium. In the EIT scheme, the presence of the control field produces the so-called dark atomic state. As a consequence, absorption is avoided and the medium is transparent. On the contrary, in the Raman scheme, the control beam generates an off-resonance absorption peak (Raman absorption): the medium is absorbing.

To give a common vision of the fast storage (Raman $\pi$-pulses) and the EIT/Raman schemes, we first introduce a Loren{tz}ian susceptibility response as an archetype for absorption and its counterpart the inverted-Lorentzian that describes a generic transparency window. We will define the different terms in \ref{archetypes}.

\subsection{Three-level atoms Schr\"odinger-Maxwell model}


Following the same approach as in section \ref{sec:2PE}, the pulse propagation and storage can be modeled by the Schr\"odinger-Maxwell equations in one dimension (along $z$).  {We now give these equations for three level atoms.}

\subsubsection{Schr\"odinger equation for three-level atoms}
For three-level atoms, labeled $|g\rangle$,  $|e\rangle$ and  $|s\rangle$ for the ground, excited  and spin  states  (see fig.\ref{fig:2level_3level}, right), the rotating-wave probability amplitudes $C_g$, $C_e$ and $C_s$ respectively are governed by the time-dependent Schr\"odinger equation similar to eq.\eqref{bloch2} \cite[eq. (13.29)]{shore2011manipulating}:
\begin{align}
i \partial_t \left[
\begin{array}{c}
C_g \\
C_e \\
C_s\\
\end{array}\right]
=
\left[
\begin{array}{ccc}
0 &\displaystyle \frac{\mathcal{E}^*}{2} & 0 \\
\displaystyle\frac{\mathcal{E}}{2} & -\Delta &\displaystyle \frac{\Omega}{2} \\
0 &\displaystyle \frac{\Omega^*}{2} & - \delta \\
\end{array}\right]
\left[
\begin{array}{c}
C_g \\
C_e \\
C_s\\
\end{array}\right]
\label{bloch3}\end{align}
where $\mathcal{E}(z,t)$ and $\Omega(t)$ are  the complex envelopes of the input signal and the Raman field respectively (units of Rabi frequency).  If we consider the spin level $|s\rangle$ as empty, the Raman field is not attenuated (nor amplified) by the propagation so  $\Omega(t)$ doesn't depend on $z$. The parameters $\Delta$ and $\delta$ are the one-photon and two-photon detunings respectively  (see fig.\ref{fig:2level_3level}, right).

The atomic variables $C_g$, $C_e$ and $C_s$ depend on $z$ and $t$ for given detunings $\Delta$ and $\delta$. As in section \ref{sec:2PE}, the detunings are chosen position and  time independent. Again, decay terms can be added {\it by-hand} by introducing complex detunings for $\Delta$ and $\delta$.

\subsubsection{Maxwell propagation equation}
%

Eqs \eqref{MB_M_hom} (homogeneous ensemble) and \eqref{MB_M_inhom} (inhomogeneous) still describe the propagation of the signal  in the slowly varying envelope approximation. 
%
%
%
%
%

The two sets of equations (\ref{bloch3}\&\ref{MB_M_hom}) or (\ref{bloch3}\&\ref{MB_M_inhom}) depending if the ensemble is homogeneous or inhomogeneous are sufficient to describe the different situations that we will consider. As already mentioned in section \ref{sec:2PE}, the equations of motion can be further simplified for weak $\mathcal{E}(z,t)$  signals (perturbative regime).

\subsubsection{Perturbative regime}

The linearisation of the  Schr\"odinger-Maxwell equations (\ref{bloch3}, \ref{MB_M_hom} \&\ref{MB_M_inhom}) corresponds to the so-called perturbative regime. To the first order in perturbation, the atoms stays in the ground, $C_g \simeq 1$ because the signal is weak. The atomic evolution (eq.\ref{bloch3}) is now only given by $C_e$ and $C_s$ that we write with $\mathcal{P}\simeq C_e$ and  $\mathcal{S}\simeq C_s$ to describe the optical (polarization $\mathcal{P}$) and spin ($\mathcal{S}$) excitations \cite{GorshkovII}. The atoms dynamics from eq.\eqref{bloch3} becomes:

\begin{align}
\partial_t \mathcal{P} &= (i\Delta-\Gamma)  \mathcal{P} - i \frac{\Omega}{2} \mathcal{S} - i \frac{\mathcal{E}}{2}\label{bloch_P}\\
\partial_t \mathcal{S} &=  - i \frac{\Omega^*}{2}  \mathcal{P} + i \delta \mathcal{S} \label{bloch_S}
\end{align}

We have introduced the optical homogeneous linewidth $\Gamma$ that will be used later. The decay of the spin is neglected which would correspond to an infinite storage time when the excitation in shelved into the spin coherence. This is an ideal case.

The Raman field $\Omega(t)$ is unaffected by the propagation if the spin state is empty. The Raman pulse keeps its initial temporal shape so there is  no differential propagation equation governing $\Omega(t)$. This a major simplification especially when a numerical integration (along $z$) is necessary.
We will only consider real envelope  $\Omega(t)$ for the Raman field. Nevertheless, a complex envelope can still be used if the Raman field is chirped for example \cite{minar_chirped}.
The exact same set of equations can alternatively be derived from the density matrix formalism in the perturbative regime, the terms $\mathcal{P}$ and $\mathcal{S}$ representing the off-diagonal coherences of the $|g\rangle$-$|e\rangle$ and  $|g\rangle$-$|s\rangle$ transitions respectively owing to the $C_g\simeq1$ hypothesis.

Using the polarization $ \mathcal{P}(t,\Delta)$, the Maxwell equations \eqref{MB_M_hom} and \eqref{MB_M_inhom} are rewritten as:
\begin{equation}
\partial_z\mathcal{E}(z,t)+ \frac{1}{c}\partial_t\mathcal{E}(z,t)=
-\displaystyle{i \alpha} \Gamma \mathcal{P}(t) \label{MB_M_hom_pert}
\end{equation}
 
or for inhomogeneous ensembles as:
\begin{equation}
\partial_z\mathcal{E}(z,t)+ \frac{1}{c}\partial_t\mathcal{E}(z,t)=
-\displaystyle\frac{i \alpha}{\pi}\int_\Delta g\left(\Delta\right)  \mathcal{P}(t,\Delta) d\Delta \label{MB_M_inhom_pert}
\end{equation}

This formalism is sufficient to describe the different situations we will consider now. The simplified perturbative set of coupled equations (\ref{bloch_P}\&\ref{bloch_S}) cannot be solved analytically when $\Omega(t)$ is time-varying, {thus} acting as a parametric driving. A numerical integration is usually necessary to fully recover the outgoing signal shape after the propagation given by eqs.\eqref{MB_M_hom_pert} or \eqref{MB_M_inhom_pert}. Simpler situations can still be examined to discuss the dispersive properties of a {\it slow-light} medium. When $\Omega(t)=\Omega$ is static, the susceptibility describing the linear propagation of the signal field $\mathcal{E}(z,t)$ can be explicitly derived. This is a very useful guide for the physical intuition.

\subsection{Inverted-Lorentzian and Lorentzian responses: two archetypes of {slow-light}}\label{archetypes}

Before going into details, we would like to describe qualitatively two archetypal situations without specific assumption on the underlying level structure or temporal shapes of the field. From our point of view, {\it slow-light} propagation should be considered as the precursor of storage. We use the term precursor as an allusion to the work of Brillouin \cite{brillouin} and Sommerfeld \cite{sommerfeld}.

The first situation corresponds to the well-known {\it slow-light} propagation in a transparency window. More specifically, we will assume {that} the susceptibility {is} given by an inverted-Lorentzian shape. The Lorentzian should be inverted to obtain transparency and not absorption at the center. The susceptibility is defined as the proportionality constant between the frequency dependent polarization and electric field (including the vacuum permittivity $\epsilon_0$). This latter can be directly identified from the field propagation equation as we will see later in \ref{section:TW} and \ref{section:AW}.

The second situation is the complementary. A Lorentzian (non-inverted) can also be considered to produce a retarded response. This is useful guide to described certain storage protocols and revisit the concept of {\it slow-light}. The Lorentzian response naturally comes out of the Lorentz-Lorenz model when the electron is elastically bound to the nucleus when light-matter interaction is introduced to the undergraduate students. These two archetypes represent a solid basis to interpret the different protocols we will detail in section \ref{raman_stopped} and \ref{EIT_stopped}.

\subsubsection{Transparency window of an inverted-Lorentzian}\label{section:TW}

We assume that the susceptibility {is} given by an inverted-Lorentzian. This is {the} simplest case because a group delay can be explicitly derived. Whatever is the exact physical situation, the source term on the right-hand sides of eqs \eqref{MB_M_hom_pert} or \eqref{MB_M_inhom_pert} can be replaced by a linear response in the spectral domain (linear susceptibility) when $\Omega(t)$ is static. The propagation equation would read in the spectral domain \cite[p.12]{allen2012optical}

\begin{equation} \frac{\dd\tilde{\mathcal{E}}(z,\omega)}{\dd z} +i\frac{\omega}{c}\tilde{\mathcal{E}}(z,\omega)= \displaystyle-\frac{\alpha}{2} \left[ 1- \frac{1}{1+i\omega/\Gamma_0} \right] \tilde{\mathcal{E}}(z,\omega) \label{propag_ILorentz} \end{equation}

where $\tilde{\mathcal{E}}(z,\omega)$ is the Fourier transform of $\mathcal{E}(z,t)$. The left-hand side simply describes the free-space propagation of the slowly varying envelope. The right-hand side is proportional to the inverted-Lorentzian susceptibility defining the complex propagation constant as:
\begin{equation}
\tilde{\alpha}\left(\omega\right)=-\frac{\alpha}{2} \left[ 1- \frac{1}{1+i\omega/\Gamma_0} \right]
\end{equation}

The different terms can be analyzed as follows. $ \frac{\alpha}{2} $ is the far-off resonance (or background) absorption coefficient for the amplitude $\tilde{\mathcal{E}}$ such as the intensity $|\tilde{\mathcal{E}}|^2$ decays exponentially with a coefficient $\alpha$ following the Bouguer-Beer-Lambert absorption law. The term $\displaystyle \left[ 1- \frac{1}{1+i\omega/\Gamma_0} \right]$ represents the Lorentzian shape of a transparency window (width $\Gamma_0$) that we choose as an archetype. With this definition, the susceptibility $\chi$ can be written as $\displaystyle \chi\left(\omega\right)=-\frac{2i}{k} \tilde{\alpha}\left(\omega\right)$ where $k$ is the wavevector \footnote{With our definitions, the real part of the propagation constant $\tilde{\alpha}$ gives the absorption and the imaginary part, the dispersion. For the susceptibility, this is the other way around.}.

At the center $\omega=0$, there is no absorption (complete transparency). We choose a complex Lorentzian $\displaystyle \frac{1}{1+i\omega/\Gamma_0}$ and not a real one $\displaystyle \frac{1}{1+\omega^2/\Gamma_0^2}$ because the complex Lorentzian satisfies {\it de facto} the Kramers-Kronig relation so we implicitly respect the causality. The propagation within the transparency window is given by a first-order expansion of the susceptibility when $\omega \ll \Gamma_0$ leading to

\begin{equation} \frac{\dd\tilde{\mathcal{E}}(z,\omega)}{\dd z} +i\frac{\omega}{c}\tilde{\mathcal{E}}(z,\omega) \simeq \displaystyle-\frac{\alpha}{2} i\frac{\omega}{\Gamma_0} \tilde{\mathcal{E}}(z,\omega) \end{equation}
and after integration over the propagation distance $L$
\begin{equation} \tilde{ \mathcal{E}}(L,\omega) \simeq \tilde{ \mathcal{E}}(0,\omega) \exp \left(-\frac{i \omega L }{c} \right) \exp \left(-\frac{i \omega \alpha L }{2\Gamma_0} \right) \end{equation}
or equivalently in the time domain
 \begin{equation} \label{eq:nunn2} \mathcal{E}(L,t) \simeq \mathcal{E}(0,t-\frac{L }{c}-\frac{\alpha L}{2 \Gamma_0}) \end{equation}

where $\displaystyle \frac{L }{c}+\frac{d}{2 \Gamma_0}$ is the group delay with the optical depth $d=\alpha L$. If the incoming pulse bandwidth fits the transparency window or in other words if the pulse is sufficiently long, the pulse is simply delayed by $\displaystyle \frac{d}{2 \Gamma_0}$. This latter defines the group delay.

Shorter pulses are distorted and partially absorbed when the bandwidth extends beyond the transparency window. In that case, eq.\eqref{propag_ILorentz} can be integrated analytically to give the general formal solution:

\begin{equation} \tilde{ \mathcal{E}}(L,\omega) = \tilde{ \mathcal{E}}(0,\omega) \exp \left(-\frac{i \omega L }{c} \right) \exp \left(-\frac{d}{2}  \frac{i \omega}{\Gamma_0+i\omega}\right) \label{SL_TW} \end{equation}

\begin{figure}
\centering
\fbox{\includegraphics[width=.65\linewidth]{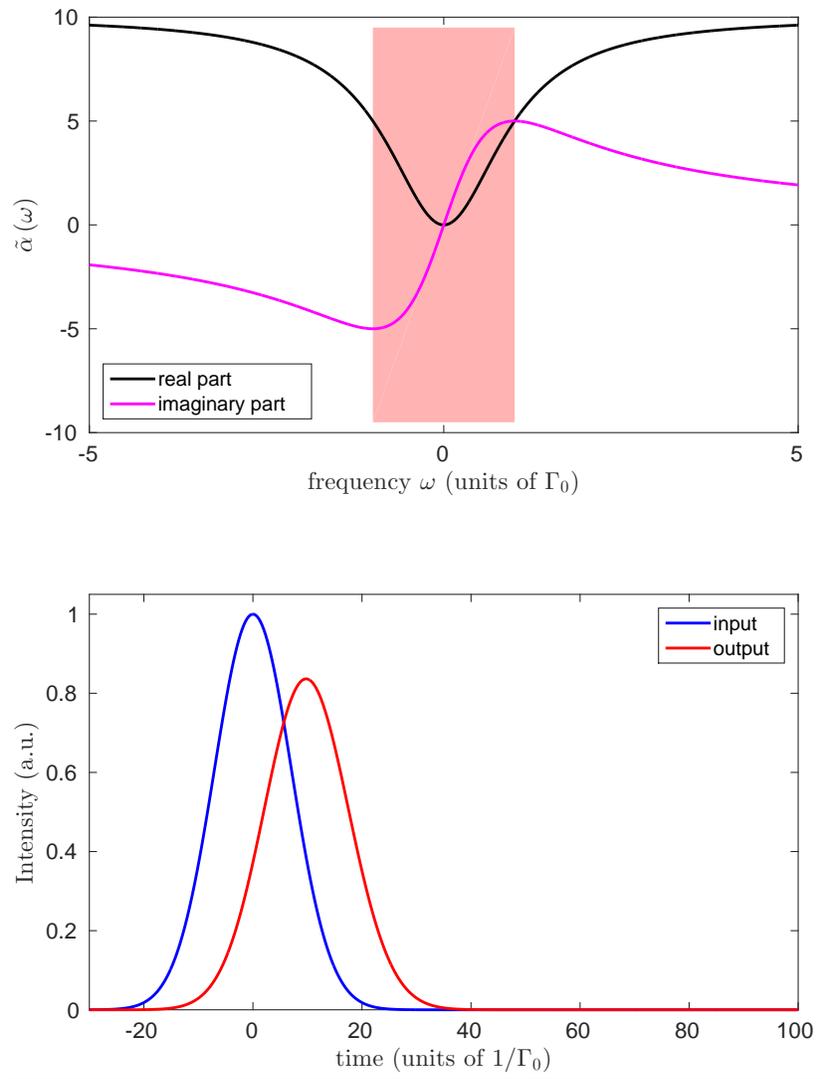}}
\caption{{\it Slow-light} in a Lorentzian transparency window. Top: real and imaginary part of the propagation constant $\tilde{\alpha}\left(\omega\right)$. The real part represents the absorption and the imaginary part the refractive index (dispersion) whose slope is the group delay. The shaded area corresponds to the {\it slow-light} region, the positive slope of the imaginary part leads to a positive group delay. Bottom: {\it Slow-light} propagation of a Gaussian incoming pulse (in blue) producing a delayed output pulse (in red) calculated from eq.\eqref{SL_TW}.}
\label{fig:SL_ILorentz} 
\end{figure}

The outgoing pulse shape $\mathcal{E}(L,t) $ is given by the inverse Fourier transform of $ \tilde{ \mathcal{E}}(L,\omega)$. As an example, we plot the outgoing pulse in fig.\ref{fig:SL_ILorentz} for a Gaussian input $\displaystyle \mathcal{E}(0,t)= \exp \left(-\frac{t^2}{2\sigma^2}\right)$. We choose $\Gamma_0=1$ and a pulse duration $\sigma =\displaystyle \frac{d}{2 \Gamma_0}$ corresponding to the expected group delay. We take $d=20$ for the optical depth, which corresponds to realistic experimental situations. 

The outgoing pulse is essentially delayed by $\displaystyle \frac{d}{2 \Gamma_0}=10$ and only weakly absorbed through the propagation. A longer pulse would lead to less absorption but the input and output would be much less separated. As we will see later, this point is critical for {\it slow-light} storage protocols.
 
\subsubsection{Dispersion of a Lorentzian}\label{section:AW}

We now consider a Lorentzian as a complementary situation. This may sound surprising for the reader familiar with the EIT transparency window. {However,} the Lorentzian is a useful reference to interpret the Raman memory {that will be discussed in section \ref{Raman}.

We {consider a} propagation constant given by 
\begin{equation}
\tilde{\alpha}\left(\omega\right)=-\frac{\alpha}{2} \frac{1}{1+i\omega/\Gamma_0} 
\end{equation}
This is a quite simple case corresponding to the transmission of an homogeneous ensemble of dipoles. To take the terminology of the previous case, one could speak of an absorption window as opposed to a transparency window. To follow up the analogy, there is no {\it slow-light} at the center of an absorption profile. The susceptibility is inverted thus leading to {\it fast-light} (negative group delay). A retarded response can still be expected but on the wings (off-resonance) of the absorption profile. As represented on fig.\ref{fig:SL_Lorentz}, the slope is negative at the center ({\it fast-light}) but it changes sign out of resonance leading to a distorted version of {\it slow-light}. Distortion are indeed expected because the dispersion cannot be considered as linear. Still, what comes out of the medium after the incoming pulse can be interpreted as a precursor for light storage.

By inverted analogy with the previous case, the propagation can be solved to the first order when the pulse bandwidth is much larger than the absorption profile (off-resonant excitation of the wings). The Lorentzian $\displaystyle \frac{1}{1+i\omega/\Gamma_0}$ simplifies to the first order in $\displaystyle \frac{\Gamma_0}{i\omega}$ leading to the solution in the spectral domain:

\begin{equation} \label{eq:propag3} \tilde{ \mathcal{E}}(L,\omega) \simeq \tilde{ \mathcal{E}}(0,\omega)  \exp \left(-\frac{i \omega L }{c} \right)  \exp \left(-\frac{\alpha L \Gamma_0}{2i \omega} \right) \end{equation}

or alternatively in the time domain

\begin{equation} \mathcal{E}(L,t) \simeq \mathcal{E}(0,t) \ast F(L,t) \label{FID_convol} \end{equation}
where $F(L,t)$ is the impulse response convoluting ($\ast$) the incoming pulse shape and analytically given by \cite{bateman1954tables}:
\begin{equation}\label{eq:FID}
F\left(L,t\right) = \delta_{t=0} - {\alpha L \Gamma_0} \frac{J_1\left(\sqrt{2d \Gamma_0 t}\right)}{\sqrt{2d \Gamma_0 t}} \mbox{ for t$>$0 and 0 elsewhere} \end{equation}

$J_1$ is the Bessel function of the first kind of order 1 with the optical depth $d=\alpha L$. $\delta_{t=0}$ is the Dirac peak. The time $\displaystyle \frac{1}{d \Gamma_0}$ appears as a typical delay due to propagation. The output shape will be distorted by the strong oscillations of the Bessel function. This can be investigated by considering the following numerical example without first order expansion. The output shape is indeed more generally given by the inverse Fourier transform of the integrated form:
\begin{equation} \tilde{ \mathcal{E}}(L,\omega) = \tilde{ \mathcal{E}}(0,\omega) \exp \left(-\frac{i \omega L }{c} \right) \exp \left(-\frac{d }{2}  \frac{\Gamma_0}{\Gamma_0+i\omega}\right) \label{SL_TA}  \end{equation}

Again we plot the outgoing pulse in fig.\ref{fig:SL_Lorentz} for a Gaussian input $\displaystyle \mathcal{E}(0,t)= \exp \left(-\frac{t^2}{2\sigma^2}\right)$ whose duration is now $\sigma =\displaystyle\frac{1}{d \Gamma_0}$ ($\Gamma_0=1$) corresponding to the expected generalized group delay. As before, the optical depth is $d=20$. 
Two lobes appear at the output (fig.\ref{fig:SL_Lorentz}) as expected from the approximated expression eq.\eqref{FID_convol} involving the oscillating Bessel function. Still, a significant part of the incoming pulse is retarded in the general sense whatever is the exact outgoing shape.

\begin{figure}
\centering
\fbox{\includegraphics[width=.65\linewidth]{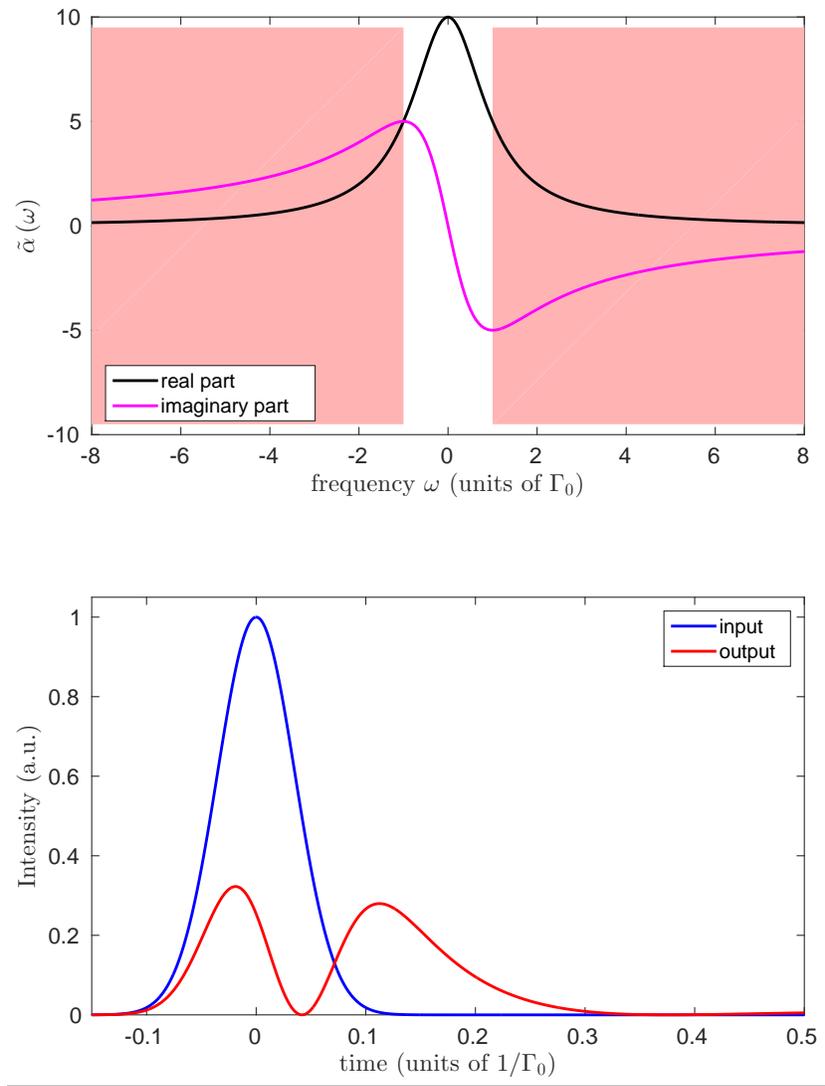}}
\caption{{\it Slow-light} from a Lorentzian absorption window. Top: real and imaginary part of the propagation constant  $\tilde{\alpha}\left(\omega\right)$. The shaded area corresponds to the {\it slow-light} region (positive group delay). Bottom: {\it Slow-light} propagation of a gaussian incoming pulse (in blue) producing a retarded output pulse (in red) calculated from eq.\eqref{SL_TA}.}
\label{fig:SL_Lorentz} 
\end{figure}

As will see now, what is retarded can be stored.

\subsubsection{A retarded response as a precursor for storage}

{\it Slow-light} is a precursor of storage called  {\it stopped-light} in that case. The transition from {\it slow} to {\it stopped-light} is summarized in fig.\ref{fig:SL_Lorentz_ILorentz_shaded}.

When input and output are well separated in time, storage is possible in principle. If we look at the standard situation of {\it slow-light} in a transparency window (fig.\ref{fig:SL_Lorentz_ILorentz_shaded}, top), we choose a frontier between input and output at half the group delay $\displaystyle \frac{d}{4 \Gamma_0}=5$. At this given moment, most of the output pulse has entered the atomic medium. There is only a small fraction of the input pulse (blue shaded area) that leaks out. This part will be lost. Concerning the output pulse, the red shaded area (subtracted from the blue area) is essentially contained inside the medium and {\it de facto} stored into the atomic excitation \cite{Shakhmuratov, ChaneliereHBSM}. The same qualitative description also applies to the retarded response from a Lorentzian absorption window (fig.\ref{fig:SL_Lorentz_ILorentz_shaded}, bottom). Storage can be expected as well but at the price of temporal shape distortion.

\begin{figure}
\centering
\fbox{\includegraphics[width=.7\linewidth]{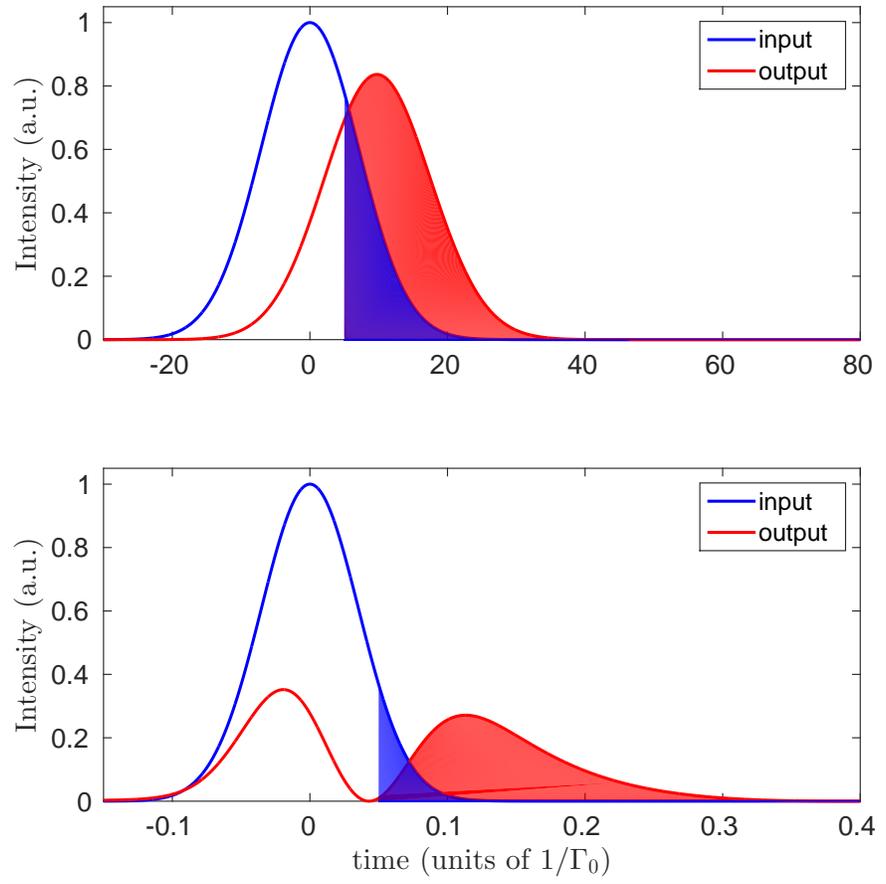}}
\caption{Top: {\it Slow-light} in a transparency window as in fig. \ref{fig:SL_ILorentz}. The shaded area after half of the group delay $\displaystyle \frac{d}{4 \Gamma_0}=5$ represents the separation between the input and the outgoing pulses. Bottom: Retarded response from a Lorentzian absorption window as in fig. \ref{fig:SL_Lorentz}. We choose for the separation between input and output the expected generalized group delay $\displaystyle\frac{1}{d \Gamma_0}=.05$.}
\label{fig:SL_Lorentz_ILorentz_shaded} 
\end{figure}

Following our interpretation, as soon as input and output are well separated, there is a moment when a f{r}action of the light is contained in the atomic excitation. This fraction defines the storage efficiency. The transition from {\it slow} to {\it stopped-light} requires to detail the specific storage protocols by giving a physical reality to the (inverted-)Lorenzian susceptibility. {\it Slow-light} ensures that the optical excitation is transiently contained in the atomic medium. For {permanent} storage {and on-demand readout}, it is necessary to act dynamically on the atomic excitation as we will see now. More precisely, the shelving of the excitation into the spin (by a brief Raman $\pi$-pulses or by switching off the control field as we will see in \ref{raman_stopped} and \ref{EIT_stopped} respectively) prevents the radiation of the retarded response. The excitation is trapped in the atomic ensemble. The evolution is resumed at the retrieval stage  by the reversed operation (by a second brief Raman pulses or by switching on the control field).

Before going into details of the storage schemes, we briefly show that the correct orders of magnitude for the efficiencies can be derived from our simplistic vision. From fig.\ref{fig:SL_Lorentz_ILorentz_shaded}, we can roughly evaluate the efficiency by subtracting the blue from the red area assuming the incoming energy (integral of the incoming pulse) is one. We find for {\it slow-light} in a transparency window (inverted-Lorentzian profile) a potential efficiency of  43\% and for the retarded response of an absorption window (Lorentzian profile) 32\%.

We will keep these numbers as points of comparison for specific protocols that we will first explicitly connect to the {\it slow-light} propagation from an inverted-Lorentzian or a Lorentzian and then numerically simulate with the previously established Schr\"odinger-Maxwell equations.

\subsection{Fast storage and retrieval with brief Raman $\pi$-pulses}\label{raman_stopped}

Our approach is based on the fact that {\it slow-light} is associated with the transient storage of the incoming pulse into the atomic excitation. A simple method to store {more} permanently the excitation is to convert instantaneously the optical excitation into a spin wave. This can be done by a Raman $\pi$-pulse as proposed in different protocols. We will now go into details and properly define the level structure and the temporal sequence required to implement the previously discussed archetypes (sections \ref{section:TW} and \ref{section:AW}). We will consider two specific protocols: the spectral hole memory and the free induction decay memory proposed in \cite{Lauro1} and \cite{Vivoli} respectively.

\subsubsection{Spectral hole memory}\label{SHOME}

The spectral hole memory has been proposed by Lauro {\it et al.} in \cite{Lauro1}  and partially investigated experimentally in \cite{Lauro2}. The protocol has been successfully implemented in \cite{SHOME} at the single photon level with a quite promising efficiency of 31\%. An inhomogeneously broaden ensemble is first considered. A spectral hole is then burnt into the inhomogeneous distribution. This situation is realistic and corresponds to rare-earth doped crystals for which the spectral hole burning mechanism, as spectroscopic tool, can be efficiently used to sculpt the absorption profile \cite{liu2006spectroscopic}. When the hole profile is Lorenztian, the propagation of a weak signal pulse precisely corresponds to the situation \ref{section:TW} as we will see now.

The atomic evolution is described by eqs.(\ref{bloch_P}\&\ref{bloch_S}) and the propagation by eq.\eqref{MB_M_inhom_pert}. The signal $\mathcal{E}(z,t)$ propagates initially through the atomic distribution described by 

\begin{equation}
g\left(\Delta\right)=\left[ 1- \frac{1}{1+\left(\Delta/\Gamma_0\right)^2} \right]\label{g_shome}
\end{equation}
 where $\Gamma_0$ is the spectral hole width.

The Raman field is initially off and is only applied for the rapid conversion into the spin wave. When the Raman field is off, the evolution eq.\eqref{bloch_P} reads as $\displaystyle \partial_t \mathcal{P} = (i\Delta-\Gamma)  \mathcal{P}  - i \frac{\mathcal{E}}{2}$. The coherence lifetime $1/\Gamma$ (inverse of the homogeneous linewidth) is assumed to be much longer than the time of the experiment such as in the spectral domain we write in the limit $\Gamma \rightarrow 0$
\begin{equation} \tilde{\mathcal{P}}(\Delta,\omega)= \displaystyle \frac{ \tilde{\mathcal{E}}(z,\omega)}{2\left(\Delta-\omega\right)} \end{equation}

So the propagation reads as

\begin{equation} \frac{\dd\tilde{\mathcal{E}}(z,\omega)}{\dd z} +i\frac{\omega}{c}\tilde{\mathcal{E}}(z,\omega)= \displaystyle-\frac{\alpha}{2} \tilde{\mathcal{E}}(z,\omega) \frac{i}{\pi} \int_\Delta \frac{g\left(\Delta\right)}{\Delta-\omega}  d\Delta  \end{equation}

The term $\displaystyle \frac{i}{\pi} \int_\Delta \frac{g\left(\Delta\right)}{\Delta-\omega}  d\Delta$ represents the susceptibility. The integral over $\Delta$ ensures that the Kramers-Kroning relations are satisfied. This last term is then given by the Hilbert transform of the distribution $g\left(\Delta\right)$ so we have $\displaystyle \frac{i}{\pi} \int_\Delta \frac{g\left(\Delta\right)}{\Delta-\omega}  d\Delta =\left[ 1- \frac{1}{1+i\omega/\Gamma_0} \right]$. The propagation of the signal is indeed given by eq.\eqref{propag_ILorentz} as described in section \ref{section:TW} and as represented in figs.\ref{fig:SL_ILorentz} and \ref{fig:SL_Lorentz_ILorentz_shaded} (top). The delayed pulse (or at least the fraction which is sufficiently separated from the input) can be stored as represented by shaded areas in fig.\ref{fig:SL_Lorentz_ILorentz_shaded} (top). As proposed in  \cite{Lauro1}, a Raman $\pi$-pulse can be used to shelve the optical excitation into the spin. A second  Raman $\pi$-pulse triggers the retrieval. They are applied on resonance ($|s\rangle$-$|e\rangle$ transition) so $\delta=0$ in eq.\eqref{bloch_S}.

When the input and the output overlap as in many realistic situations or in other words when the signal cannot be fully compressed spatially into the medium, the storage step cannot be solved analytically.  A numerical simulation of the Schr\"odinger-Maxwell equations  is necessary (eqs.\ref{bloch_P}\&\ref{bloch_S} with $\Gamma=0$ and $\delta=0$ and eq.\eqref{MB_M_inhom_pert} for the propagation). For a given inhomogeneous detuning  $\Delta$, we calculate the atomic evolution eqs.(\ref{bloch_P}\&\ref{bloch_S})  by using a fourth-order Runge-Kutta method. After summing over the inhomogeneous broadening, the output pulse is given by  integrating eq.\eqref{MB_M_inhom_pert} along $z$ using the Euler method. A good test for the numerical simulation is to calculate the output pulse without Raman pulses and compare it to the analytic expression from the Fourier transform of eq.\eqref{SL_TW}.

The Raman $\pi$-pulses defined by $\Omega(t)$ are taken as two Gaussian pulses whose area is $\pi$. Following the insight of  fig.\ref{fig:SL_Lorentz_ILorentz_shaded} (top), we choose to apply the first Raman pulse at half the group delay $\displaystyle \frac{d}{4 \Gamma_0}=5$. The second Raman pulse is applied later to trigger the retrieval. The result of the storage and retrieval sequence is presented in fig.\ref{fig:plot_outputIO_SHOME}.

\begin{figure}
\centering
\fbox{\includegraphics[width=.8\linewidth]{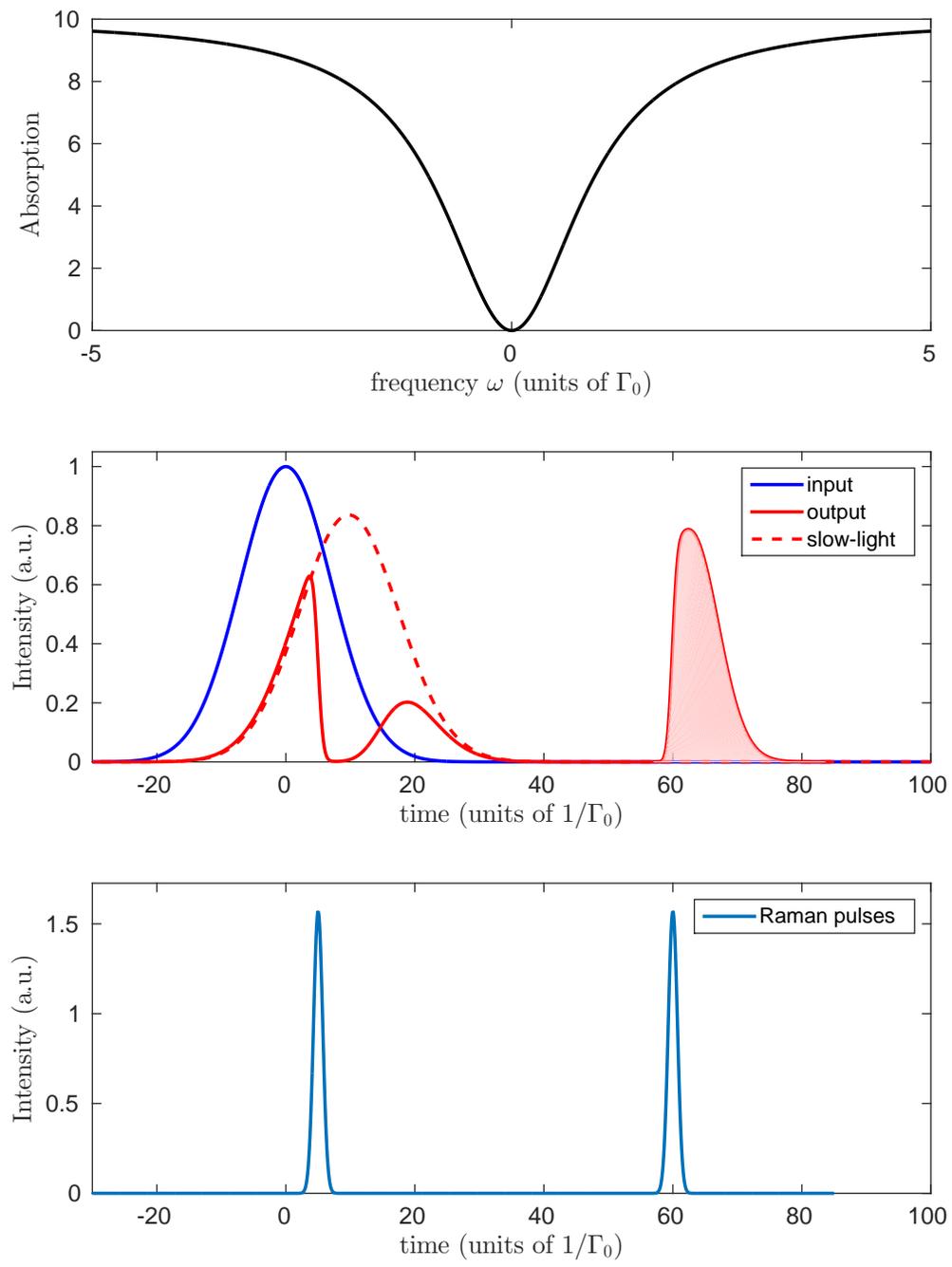}}
\caption{Spectral hole memory protocol. Top: Absorption profile from the inhomogeneous distribution $g$ defined by eq.\eqref{g_shome}. Middle: Incoming signal (in blue) and outgoing stored pulse (in red). We have also represented the {\it slow-light} pulse (dashed red) as a reference when there is no Raman pulse. Bottom: Two Raman $\pi$-pulses. The first one is applied at half the group delay $\displaystyle \frac{d}{4 \Gamma_0}=5$ and the second later on to trigger the retrieval.
}
\label{fig:plot_outputIO_SHOME}
\end{figure}

As parameters for the simulation, we choose the same as in \ref{section:TW} meaning $\Gamma_0=1$, an optical depth of $d=20$ and $\sigma =\displaystyle \frac{d}{2 \Gamma_0}=10$ for the incoming pulse duration. The Raman pulses should be ideally short to uniformly cover the signal excitation bandwidth. In our case, we choose Gaussian Raman pulses with a duration $\sigma_\pi=1$ (ten times shorter than the signal).

In fig.\ref{fig:plot_outputIO_SHOME} (middle), we clearly see that the first Raman pulse somehow clips the {\it slow-light} pulse corresponding to the shelving of the optical excitation into the spin wave. At this moment, since part of the input pulse is still present, a small replica is generated leaving the medium at time $20$ in our units. The second Raman $\pi$-pulse (at time $60$) triggers the retrieval that we shaded in pale red. A realistic storage situation cannot be fully described by our qualitative picture in fig.\ref{fig:SL_Lorentz_ILorentz_shaded} where the {\it slow-light} signal would be clipped, frozen, delayed and retrieved later on. The complex propagation of clipped Gaussian excitations in the medium can only be accurately embraced by a numerical simulation. The naive picture gives nonetheless a qualitative guideline to understand the storage. A quantitative analysis can be performed by evaluating the stored energy corresponding to the pale red shaded area under the intensity curve. From the simulation, we obtain 36\% to be compared with the 43\% obtained from fig.\ref{fig:SL_Lorentz_ILorentz_shaded}. The agreement is satisfying given the numerical uncertainties and the complexity of the propagation process when the Raman pulses are applied. We now turn to the complementary situation described in section \ref{section:AW} by following the same procedure.

\subsubsection{Free induction decay memory} \label{FID}

Free induction decay memory to take the terminology of the article by Caprara Vivoli {\it et al.} \cite{Vivoli} has not been yet implemented in practice despite a connection with the extensively studied {\it slow-light} protocols. The situation actually corresponds to our description in section \ref{section:AW} where the response of a Lorenztian to a pulsed excitation is considered. This response has been analyzed as a generalization of the free induction decay phenomenon (FID) by Caprara Vivoli {\it et al.} \cite{Vivoli}. The FID is usually observed in low absorption sample after a brief excitation. The analysis in terms of FID is perfectly valid. The response that we considered with eq.\eqref{eq:FID} with a first order expansion of the susceptibility falls into this framework. We analyze the same situation in different terms recovering the same reality. The excitation produces a retarded response that we consider as a generalized version of {\it slow-light}. This semantically connects {\it slow-light} and FID in the context of optical storage.

For the FID memory, the transition can be inhomogeneously or homogeneously broaden{ed}. Both lead to the same susceptibility. We assume the medium homogenous with a linewidth $\Gamma$ thus simplifying the analysis and the numerical simulation, the propagation being given by eq.\eqref{MB_M_hom_pert}.

As in the spectral hole memory, the Raman field is initially off and serves as a rapid conversion into the spin wave by the application of a $\pi$-pulse. When the Raman field is off, the evolution (eq.\ref{bloch_P}) reads as $\displaystyle \partial_t \mathcal{P} = -\Gamma  \mathcal{P}  - i \frac{\mathcal{E}}{2}$. The signal is directly applied on resonance so $\Delta=0$. We then obtain for the polarization \begin{equation} \tilde{\mathcal{P}}(\Delta,\omega)= \displaystyle \frac{ -i \tilde{\mathcal{E}}(z,\omega)}{2\left(i\omega+\Gamma\right)} \end{equation}
 and the propagation
 
 \begin{equation} \frac{\dd\tilde{\mathcal{E}}(z,\omega)}{\dd z} +i\frac{\omega}{c}\tilde{\mathcal{E}}(z,\omega)= \displaystyle-\frac{\alpha}{2} \tilde{\mathcal{E}}(z,\omega)  \frac{\Gamma}{i\omega+\Gamma}   \end{equation} whose solution is indeed given by eq.\eqref{SL_TA}. The output pulse is distorted and globally affected by a typical delay $\displaystyle \frac{1}{d \Gamma}$. A first Raman $\pi$-pulse can be applied at this moment. A second  Raman $\pi$-pulse triggers the retrieval. As in the spectral hole memory (section \ref{SHOME}), they are applied on resonance so $\delta=0$ in eq.\eqref{bloch_S}.
 
The complete protocol (when Raman pulses are applied) can only be simulated numerically from the Schr\"odinger-Maxwell equations  (eqs.\ref{bloch_P}\&\ref{bloch_S} with $\Delta=0$ and $\delta=0$ and eq.\eqref{MB_M_hom_pert} for the propagation in an homogeneous sample). Following fig.\ref{fig:SL_Lorentz_ILorentz_shaded} (bottom), we choose to apply the first Raman pulse at the generalized group delay $\displaystyle \frac{1}{d \Gamma}$. The second Raman pulse is applied later on to trigger the retrieval.

For the simulation, we again choose the parameters used in section \ref{section:AW} namely a linewdith $\Gamma=1$  and a signal pulse duration of $\sigma =\displaystyle\frac{1}{d \Gamma}=0.05$  corresponding to the expected generalized group delay for an optical depth $d=20$. The Raman pulses have a duration $\sigma_\pi=0.005$ (ten times shorter than the signal) and a $\pi$-area. The result is presented in fig.\ref{fig:plot_outputIO_FID}.

\begin{figure}
\centering
\fbox{\includegraphics[width=.8\linewidth]{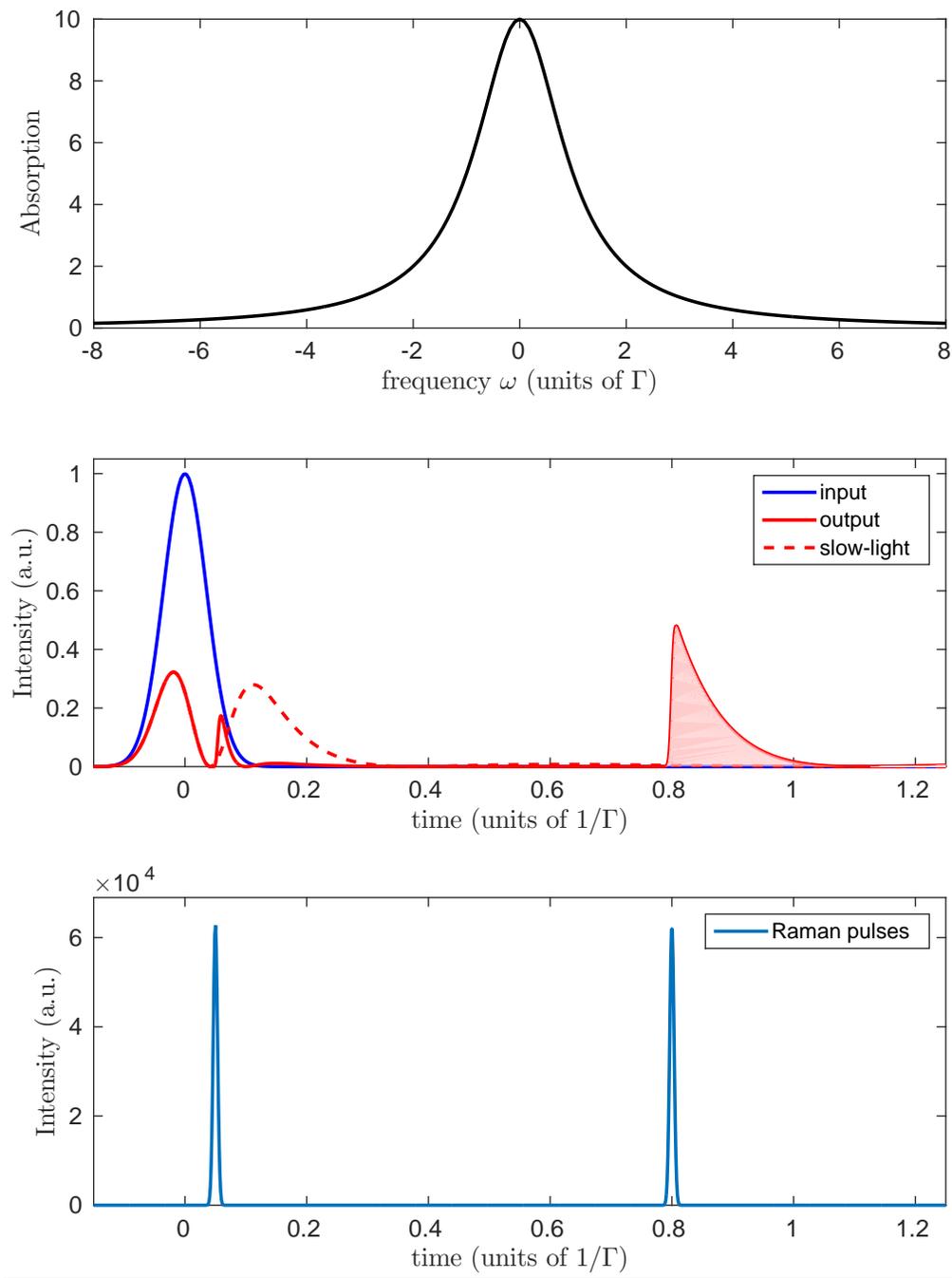}}
\caption{Free induction decay memory.  Top: Lorentzian absorption profile from an homogeneous sample. Middle: Incoming signal (in blue) and outgoing stored pulse (in red). We have also represented the {\it slow-light} pulse (dashed red) as a reference when there is no Raman pulse. Bottom: Two Raman $\pi$-pulses. The first one is applied at $\displaystyle \frac{1}{d \Gamma}=0.05$ and the second later on to trigger the retrieval.}
\label{fig:plot_outputIO_FID} 
\end{figure}

We retrieve the tendencies of the spectral hole memory. The first Raman pulse clips the {\it slow-light} pulse by storing the excitation into the spin state. As opposed to the propagation in the spectral hole, there is no replica after the first Raman pulse. This replica is strongly attenuated (slightly visible in fig.\ref{fig:plot_outputIO_FID})  because it propagates through the absorption window. We trigger the retrieval at time $0.8$ by a second Raman $\pi$-pulse. The temporal output shape cannot be compared to a clipped version of the input or the {\it slow-light} pulse. This situation is clearly more complex than the spectral hole memory. That being said, the resemblance of the output shape with a exponential decay somehow a posteriori justifies the term FID for this memory scheme. The red pale shaded area represents an efficiency of 42\% with respect to the input pulse energy. This numerical result has to be compared with 32\% obtained from fig.\ref{fig:SL_Lorentz_ILorentz_shaded}. The agreement is not satisfying even if it is difficult to have a clear physical vision of the pulse distortion induced by the  propagation at large optical depth. The order of magnitude is nevertheless correct.

The FID protocol can be optimally implemented by using an exponential rising pulse for the incoming signal (instead of a Gaussian in  fig.\ref{fig:plot_outputIO_FID}, middle) as analyzed in the reference paper \cite{Vivoli}. In that case, input (rising exponential) and output (decaying exponential) pulse shapes are time-reversed corresponding to the optimization procedures defined in \cite{GorshkovII, GorshkovPRL} and implemented in the EIT/Raman memories \cite{Novikova, nunnMultimode, zhou2012optimal}


Starting from two representative situations in \ref{section:TW} and \ref{section:AW} where the dispersion produces a retarded response from the medium, we have analyzed two related protocols in \ref{SHOME} and \ref{FID} that qualitatively corresponds to the storage of this delayed response. Except in a recent implementation \cite{SHOME}, these protocols have not been much considered in practice despite a clear connection with the archetypal propagation through the Lorentzian susceptibility of an atomic medium. On the contrary, electromagnetically induced transparency and Raman schemes are well-known and extensively studied experimentally. We will show now that they follow the exact same classification thus enriching our comparative analysis.

\subsection{Electromagnetically induced transparency and Raman schemes}\label{EIT_stopped}

Starting from two pioneer realizations \cite{phillips2001storage, liu2001observation}, the implementation of the electromagnetically induced transparency (EIT) scheme has been continuously active in the prospect of quantum storage. As opposed to the spectral hole (section \ref{SHOME}) and the free induction decay (section \ref{FID}) memories and recalling to the reader the main difference, EIT is not based on the transient excitation of the optical transition that is rapidly transfered into the spin by a Raman $\pi$-pulse. In EIT, the direct optical excitation is avoided by precisely using the so-called dark state in a $\Lambda$-system \cite{fleischhauer2000dark, FLEISCHHAUER2000395}. Practically, a control field is initially applied on the Raman transition to obtain {\it slow-light} from the $\Lambda$-system susceptibility\footnote{Inversely, for the spectral hole in \ref{SHOME} and the free induction decay in \ref{FID} memory, the Raman field is initially off.}. As a first cousin, the Raman memory scheme has been proposed and realized afterward \cite{nunnPRA,nunnNat}. EIT and Raman memories are structurally related by a common $\Lambda$-system which is weakly excited by the signal on one branch and controlled by a strong laser on the Raman branch (see fig.\ref{fig:2level_3level}). The main difference comes from the excited state detuning. For EIT scheme, the control field is on resonance. For the Raman scheme, the control field is off resonance. As we will see now, these two situations actually corresponds to the archetypal dispersive profiles described in \ref{section:TW} and \ref{section:AW} respectively.

\subsubsection{Electromagnetically induced transparency memory}\label{EIT}

The atomic susceptibility in a $\Lambda$-system is derived from eqs.(\ref{bloch_P}\&\ref{bloch_S}). Initially, the Raman field $\Omega$ is on and assumed constant in time. In EIT, the signal and control fields are on resonance so $\Delta=\delta=0$. The medium is assumed homogeneous even if the calculation can be extended to the inhomogeneously broaden systems \cite{kuznetsova2002atomic}. The propagation is here given by eq.\eqref{MB_M_hom_pert}.
 
In the spectral domain, eqs.(\ref{bloch_P}\&\ref{bloch_S}) read as 
\begin{equation} \tilde{\mathcal{P}}(\Delta,\omega)= \displaystyle \frac{ -i \tilde{\mathcal{E}}(z,\omega)}{2\left(i\omega+\Gamma-\displaystyle i\frac{\Omega^2}{4\omega}\right)} \end{equation}

We have assumed the control field {to be} real so the intensity is written as $\Omega^2$ which can be generalized to $\Omega^*\Omega$ for complex values (chirped Raman pulses for example).

The linear susceptibility for the signal field is defined by the propagation equation in the spectral domain
 \begin{equation} \frac{\dd\tilde{\mathcal{E}}(z,\omega)}{\dd z} +i\frac{\omega}{c}\tilde{\mathcal{E}}(z,\omega)= \displaystyle-\frac{\alpha}{2} \tilde{\mathcal{E}}(z,\omega)  \frac{\Gamma}{i\omega+\Gamma - \displaystyle i\frac{\Omega^2}{4\omega} }   \label{propag_EIT}\end{equation}

The term $i \displaystyle  \frac{\Omega^2}{4\omega}$ induces the transparency when the control field is applied. Without control, the susceptibility is Lorentzian and the signal would be absorbed following the Bouguer-Beer-Lambert absorption law (eq.\ref{bouguer}). On the contrary, when the control field is on, the susceptibility is zero when $\omega \rightarrow 0$. This corresponds to the resonance condition because we assumed  $\Delta=\delta=0$. The analysis can be further simplified by considering a first order expansion within the transparency window.

The width of the transparency window is $\gammaeit=\displaystyle \frac{\Omega^2}{4\Gamma}$ which is usually much narrower than $\Gamma$. So, in the limit $\omega \ll \gammaeit \ll \Gamma$, the propagation constant reads as
 \begin{equation} \displaystyle-\frac{\alpha}{2} \frac{\Gamma}{i\omega+\Gamma - \displaystyle i\frac{\Omega^2}{4\omega} } \simeq  \displaystyle-\frac{\alpha}{2} \left[ 1- \frac{1}{1+i\omega/\gammaeit} \right]   \end{equation}

The EIT window is locally an inverted-Lorentzian that we have analyzed in \ref{section:TW}. The {\it slow-light} propagation is precisely due to the presence of the control field. The so-called dark state corresponds to a direct spin wave excitation whose radiation is mediated by the control field. The storage simply requires the extinction of the control field. The excitation is then frozen in the non-radiating Raman coherence because of the absence of control. The retrieval is triggered by switching the control back on.

The {\it stopped-light} experimental sequence can be simulated numerically from eqs.(\ref{bloch_P}\&\ref{bloch_S}) and eq.\eqref{MB_M_hom_pert}. For the parameters, we choose the same as in \ref{section:TW} and \ref{SHOME}, meaning $\gammaeit=1$ so the width of the inverted-Lorentzian is $1$. We opt for $\Omega=4$ and $\Gamma=4$ so the condition $\gammaeit \ll \Gamma$ is vaguely satisfied. Again the  optical depth is $d=20$ and $\sigma =\displaystyle \frac{d}{2 \gammaeit}=10$ is the incoming pulse duration. At time $\displaystyle \frac{d}{4\gammaeit}=5$, half the group delay, the control field is switched off ($\Omega=0$). The result is plotted in fig.\ref{fig:plot_outputIO_EIT}.

\begin{figure}
\centering
\fbox{\includegraphics[width=.8\linewidth]{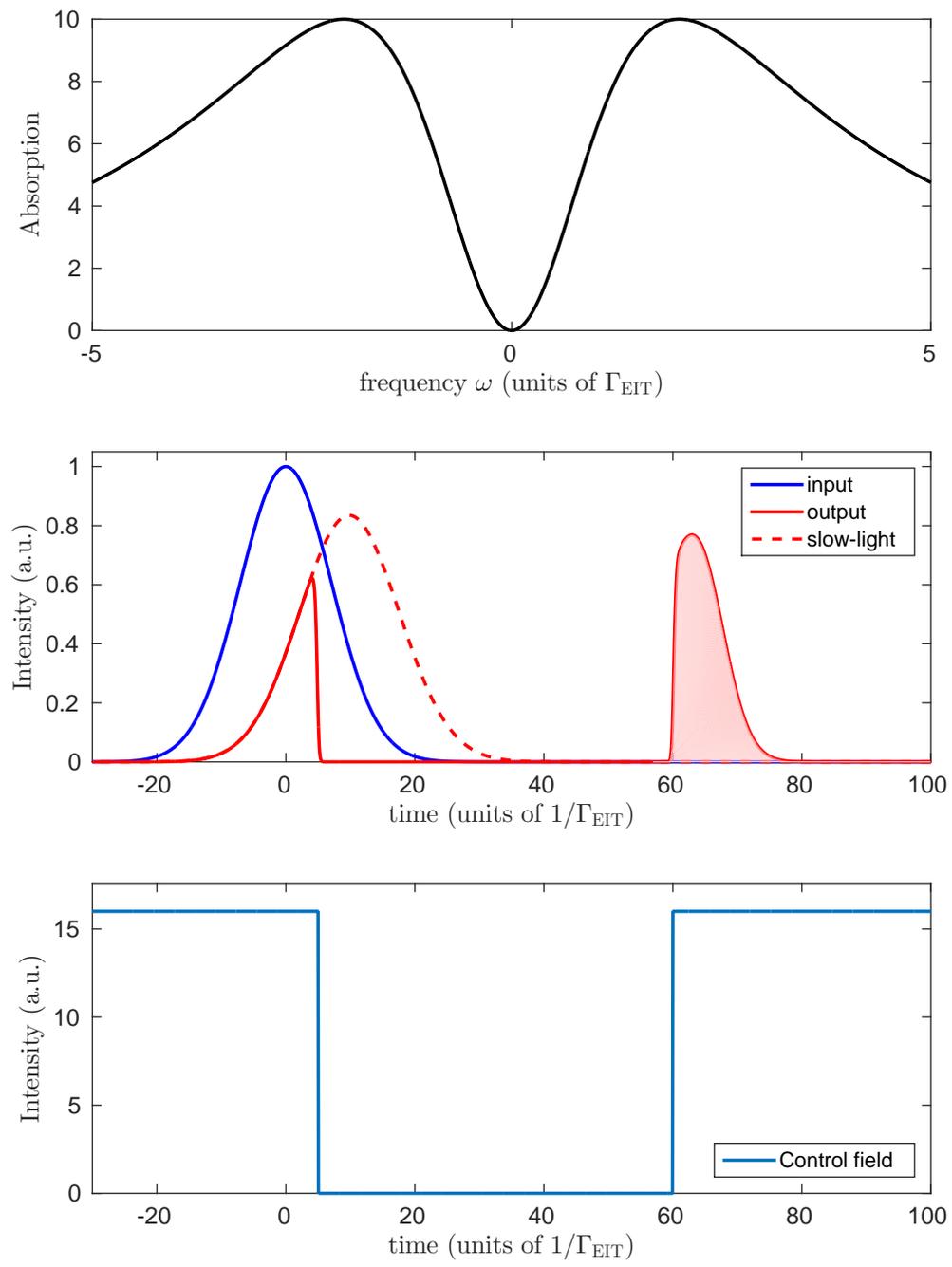}}
\caption{Electromagnetically induced transparency memory.  Top: EIT absorption profile. Middle: Incoming signal (in blue) and outgoing stored pulse (in red). We have also represented the {\it slow-light} pulse (dashed red) as a reference from eq.\eqref{propag_EIT} when the control field is always on. Bottom: The control is initially on with $\Omega^2=16$. It is switched off at half of the group delay $\displaystyle \frac{d}{4 \gammaeit}=5$ and back on later to trigger the retrieval.}
\label{fig:plot_outputIO_EIT} 
\end{figure}

Although the condition $\gammaeit=1 \ll \Gamma=4$ is only roughly satisfied so the absorption profile is not a pure inverted-Lorenzian, this has a minor influence on the {\it slow} and {\it stopped-light} pulses. The resemblance with fig.\ref{fig:plot_outputIO_SHOME} is striking even if the spectral hole and EIT  memories cover different physical realities. From fig.\ref{fig:plot_outputIO_EIT}, we can estimate the efficiency (red pale area) to 42\% thus retrieving the same expected efficiency as the spectral hole memory. One difference between fig.\ref{fig:plot_outputIO_SHOME}(middle) and fig.\ref{fig:plot_outputIO_EIT}(middle) is worth being commented: there is not {\it slow-light} replica after the time $\displaystyle \frac{d}{4 \gammaeit}=5$ for the EIT situation. This replica is absorbed in that case because when the control field is switched off, the absorption is fully restored. The presence or the absence of replicas does not change the efficiency because they correspond to a fraction of the incoming pulse that is not compressed in the medium. This leaks out and is lost anyway.

We will now complete our picture by considering the Raman memory and emphasize the resemblance with the free induction decay discussed in \ref{FID}.

\subsubsection{Raman memory}\label{Raman}

The Raman memory scheme is based on the same $\Lambda$-structure when a control field is applied far off-resonance on the Raman branch \cite{nunnPRA,nunnNat, sheremet} (see fig.\ref{fig:2level_3level}). The condition $\Delta \gg \Gamma$ defines literally the Raman condition as opposed to EIT where the control is on resonance ($\Delta=0$). The absorption profile exhibits the so-called Raman absorption peak. This Lorentzian profile is the basis for a retarded  response that we introduced in \ref{section:AW}. We first verify that the far off-resonance excitation of the control leads to a Lorenztian susceptibility for the signal. As in the EIT case (see \ref{EIT}), the atomic evolution in a $\Lambda$-system is given by eqs.(\ref{bloch_P}\&\ref{bloch_S}) and the propagation by eq.\eqref{MB_M_hom_pert}. 
The polarization is 
\begin{equation} \tilde{\mathcal{P}}(\Delta,\omega)= \displaystyle \frac{ -i \tilde{\mathcal{E}}(z,\omega)}{2\left(i\omega-i\Delta+\Gamma-\displaystyle i\frac{\Omega^2}{4(\omega-\delta)}\right)} \end{equation}

The two-photon detuning $\delta$ is not zero in that case because the Raman absorption peak in shifted by the AC-Stark shift (light shift). The signal pulse has to be detuned by $\delta=\displaystyle \frac{\Omega^2}{4\Delta}$, the light-shift, to be centered on the Raman absorption peak. Following the same approach as in the EIT case, the analysis can be simplified by a first order expansion is $\omega$. Assuming the incoming pulse bandwidth $\omega$ smaller than the light shift $\delta,$ the latter being smaller than the detuning $\Delta$, that is $\omega \ll \delta \ll \Delta$, the propagation constant reads to the first order in $\omega$ as
 \begin{equation}
 \displaystyle-\frac{\alpha}{2} \frac{\Gamma}{i\omega-i\Delta+\Gamma - \displaystyle i\frac{\Omega^2}{4(\omega-\delta)} } \simeq  \displaystyle-\frac{\alpha}{2} \frac{\gammar}{\gammar+i\omega}  \label{Raman_susceptibility}
\end{equation}
 where $\displaystyle \gammar=\frac{\Omega^2\Gamma}{4\Delta^2}$ is the width of the Raman absorption profile. This Lorentzian absorption profile can be used for storage as discussed in \ref{section:AW} and \ref{FID}. As in the EIT case, the storage is triggered by the extinction of the Raman control field. To fully exploit the analogy with \ref{section:AW} and \ref{FID}, we will choose $\gammar=1$. To satisfy the far off resonance Raman condition, we choose $\Gamma=10$ and $\Delta=1000$ thus imposing $\Omega=200\sqrt{10}$ and $\delta=100$. We run a numerical simulation of eqs.(\ref{bloch_P}\&\ref{bloch_S})  and eq.\eqref{MB_M_hom_pert} with a Gaussian incoming pulse whose duration is again $\sigma =\displaystyle\frac{1}{d \Gamma_R}=0.05$  and with an optical depth $d=20$. The result is presented in fig.\ref{fig:plot_outputIO_Raman} where the control Raman field is switched off at time $\displaystyle\frac{1}{d \Gamma_R}=0.05$ (the typical delay) and switched back on later to trigger the retrieval. 

\begin{figure}
\centering
\fbox{\includegraphics[width=.8\linewidth]{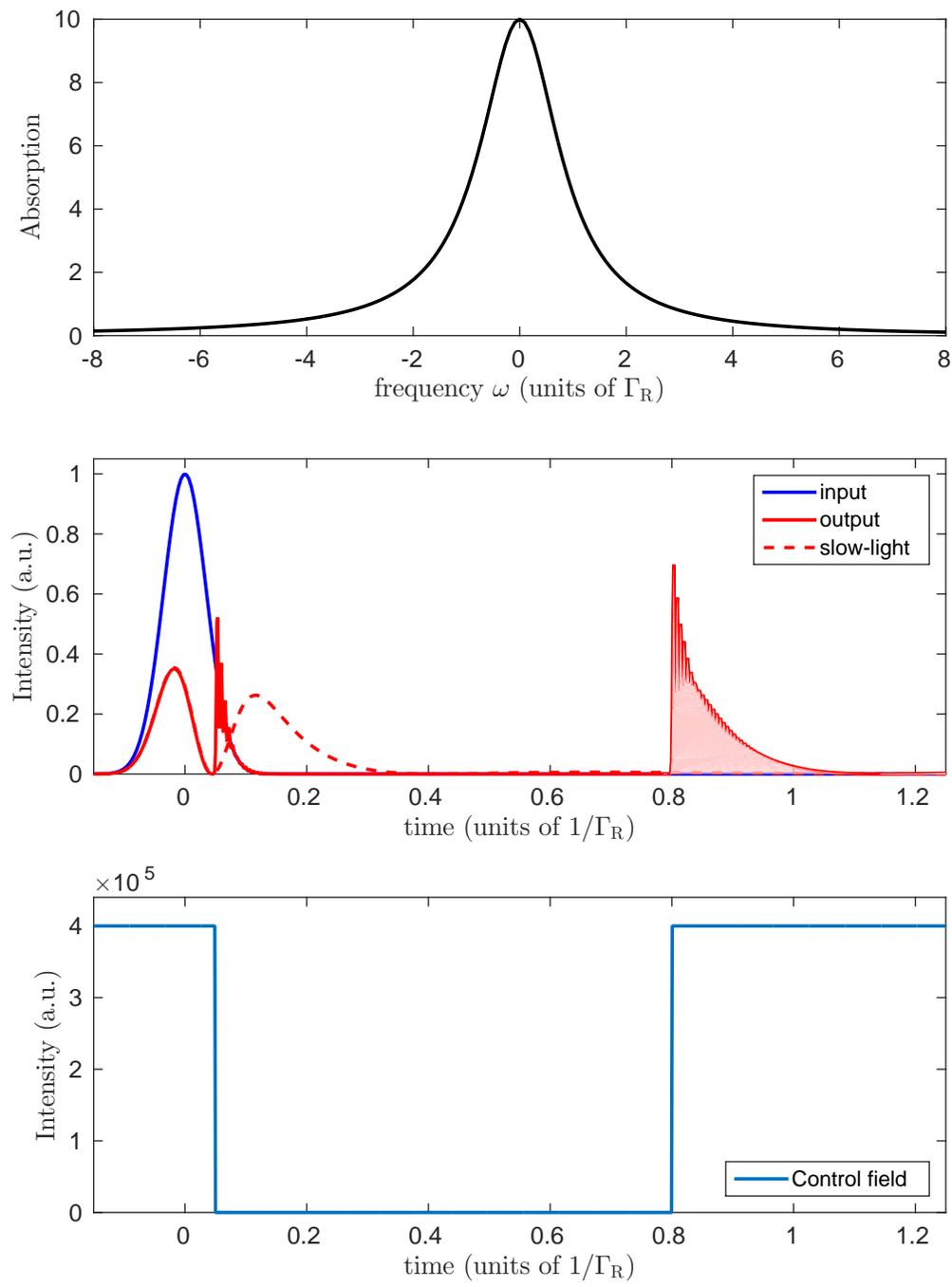}}
\caption{Raman memory.  Top: Raman absorption profile. Middle: Incoming signal (in blue) and outgoing stored pulse (in red). We have also represented the {\it slow-light} pulsed (dashed red) as a reference from eq. \eqref{Raman_susceptibility} when the Raman control field is always on. Bottom: The control is initially on with $\Omega=200\sqrt{10}$ and is switched later on to trigger the retrieval.}
\label{fig:plot_outputIO_Raman} 
\end{figure}

The resemblance with fig.\ref{fig:plot_outputIO_FID} is noticeable. Transient rapid oscillations appears when the control is abruptly switched, this is a  manifestation of the light-shift. Without surprise, the expected efficiency (red-pale area) is 42\% as the free induction decay memory with the same intensive parameters (see \ref{FID}).

\subsection{Summary and perspectives}

We have given in this section a unified vision of different {\it slow-light} based protocols. In this category, the ambassador is certainly the EIT scheme which has been particularly studied in the last decade with remarkable results in the quantum regime \cite{review_Ma_2017}. The linear dispersion associated with the EIT transparency window allows to define unambiguously a slow group velocity whose reduction to zero produces {\it stopped-light}. We have extended this concept to any retarded response that can be seen as a precursor for storage. This approach allows us to interpret the Raman scheme within the same framework. In that case, the group velocity cannot be defined {\it per se} but the dispersion profile still produces a retarded response that can be stored by shelving the excitation into a long lived spin state. The price to pay at the retrieval step is a significant pulse distortion even if the efficiency (input/output energy ratio) is quite satisfying. The pulse distortion at the retrieval is somehow a false problem. Distortions are more or less always present. Even in the more favorable EIT scheme, the pulse can be partially clipped because of a limited optical depth. It should be kept in mind that {quantum} repeater architectures {use} interference between outgoing photons \cite{simon2007, RevModPhys.83.33, bussieres2013prospective}. As soon as the different memories induce the same distortion, the retrieved outgoing {fields} can perfectly interfere. In that sense, the deformation can also be considered as an unitary transform between temporal modes without degrading the quantum information quality \cite{brecht2015photon, Thiel:17}.

The signal temporal deformation also raises the question of the waveform control through the storage step. We have used a simplistic model for the control field (on/off or $\pi$-pulses). A more sophisticated design of the control actually allows a build-in manipulation of the temporal and frequency modes of the stored qubit \cite{fisher2016frequency, conversion}. A quantum memory can be also considered a  versatile light-matter interface with a enhanced panel of processing functions. Waveform shaping is not considered anymore as a detrimental experimental limitation but as new degree of freedom whose first benefit is the storage efficiency \cite{Novikova, zhou2012optimal} when specific optimization procedures are implemented \cite{GorshkovPRL, GorshkovII, nunnMultimode}. The optimization strategy by temporal shaping is beyond the scope of this chapter but would certainly {deserve} a review paper by itself.

The fast storage schemes and the EIT/Raman sequences that we analyzed in parallel in sections \ref{raman_stopped} and \ref{EIT_stopped} respectively, both rely on a Raman coupling field that control the storage and retrieval steps. The fast storage schemes depend on $\pi$-pulses and the EIT/Raman sequence on a control on/off switching.  A three-level $\Lambda$-system seems to be necessary in that case.  This is not rigorously true even if  the $\Lambda$-structure is widely exploited for quantum storage. {\it Stopped-light} can indeed be obtained in two-level atoms by dynamically controlling the atomic properties \cite{simon_index, simon_dipole, chos}. Despite a lack of experimental demonstrations, these two-level alternative approaches conceptually extend the protocols away from the well-established atomic $\Lambda$-structure.

To close the loop with the previous section \ref{sec:2PE} on photon echo memories, we would like to discuss again the atomic frequency comb (AFC) protocol \cite{afc}. Despite its historical connection with the three-pulse photon echo sequence, {it has been argued that} the AFC falls in the {\it slow-light} memories \cite{afc_slow}. A judicious periodic shaping of the absorption profile, forming a comb, allows to produce an efficient echo. This latter can alternatively be interpreted as an undistorted retarded response using the terminology of section \ref{sec:SL}. This retarded part is a precursor that can be stored by shelving the excitation into spin states by a Raman $\pi$-pulse, thus definitely positioning the AFC in the fast storage schemes (section \ref{raman_stopped}).


\section{Certifying the quantum nature of light storage protocols}\label{sec:certification}

The question that we address in this section is how to prove that a storage protocol operates in the quantum regime. The most natural answer is: { by demonstrating that the quantum nature of a light beam is preserved after storage}. There are, however, several ways for a {memory to output light beams that show} quantum features.  It can simply be a light pulse, like a single photon, that cannot be described by a coherent state or a statistical mixture of coherent state \cite{glauber}.

Alternatively, a state can be qualified as being quantum when it leads to correlations between measurement results that cannot be reproduced by classical strategies based on pre-agreements and communications, as some entanglement states do. What is thus the difference between showing the capability of a given memory to store and retrieve single photons and entangled states ?

{Faithful} storage and retrieval of single photons {demonstrates} that the noise {generated by} the memory is low enough to preserve the photon statistics, even when {these statistics} cannot be reproduced by classical light. It does not show, however, that the memory preserves coherence. Furthermore, it does not prove that the memory cannot be reproduced by a classical strategy, that is, a protocol which would first {measure} the incoming photon and create another photon when requested.

{On the other hand}, the storage and retrieval of entangled states can be {implemented} to show that the memory outperforms any classical {measure}-and-prepare strategy. This is true provided that the fidelity of the storage protocol is high enough. For example, if a memory is characterized by storing one part of a two-qubit entangled state, the fidelity threshold is given by the fidelity of copies that would be created by a cloning machine taking one qubit and producing infinitely many copies. This is known to be one of the optimal {strategies} for determining an unknown qubit state \cite{Gisin1}.

Note that the fidelity reference can also be taken as the fidelity that would be obtained by a cloning protocol producing only one copy of the output state \cite{Scarani}. In this case, the goal is to {ensure} that the memory delivers the state with the highest possible fidelity, that is, if a copy exists, it cannot have a higher overlap with the input state. This {condition} is relevant whenever one wants to show the suitability of the memory for applications related to secure communications, {where third parties should not obtain information about the stored state} \cite{BB84}.

The suitability of a memory for secure communications can ultimately be certified {by} Bell tests \cite{Bell:1964kc}. In this case, the quality of the memory can be {estimated} without {assumptions} on the input state or on the measurements performed on the {retrieved} state. This {ensures} that the memory can be used in networks where secure communications can be realized over long distances with security guarantees holding independently of the details of the actual implementation.

We show in the following sections how these criteria can be tested in practice, describing separately benchmarks based on continuous and discrete variables. Various memory protocols are {used} as examples, including protocols such as {the two-pulse photon echo (2PE)} \cite{ruggiero} or the classical teleporter, which are known to be classical. In order to prove it, we first show how to compute the noise {inherent to} classical protocols by moving away from the semi-classical picture. While a fully quantized propagation model can be found in the literature \cite{GorshkovII} mirroring the semi-classical Schrödinger-Maxwell equations that we use in the previous sections, we present a toy model using an atomic chain to characterize memory protocols together with their noise (section \ref{toy_model}). Criteria are then derived first for continuous (section \ref{CV_criterion}) and then for {discrete} variables (section \ref{counting_criterion}).

}


\subsection{Atomic chain quantum model}\label{toy_model}

The aim is to derive a simple quantum model allowing to characterize different storage protocols including the noise. Although quantum, the model is very simple and uses the basic tools of quantum optics. 

\subsubsection{Jaynes-Cummings propagator}

We consider an electromagnetic field described by the bosonic operators $a$ and $a^\dag$ resonantly interacting with a single two-level atom (with levels $|g\rangle$ and $|e\rangle$) thought the Jaynes-Cummings Hamiltonian
\begin{equation}
\label{Jaynes_Cummings}
H_{\text{int}} = i \kappa (a^\dag \sigma_- - a \sigma_+).
\end{equation}
Here, $\sigma_\pm$ are atomic operators corresponding to the creation $\sigma_+ = |e\rangle\langle g| $ and annihilation $\sigma_- = |g\rangle\langle e|$ of an atomic excitation. The first term in \eqref{Jaynes_Cummings} is thus associated to the emission of a photon while the second term corresponds to its absorption. The corresponding propagator 
\begin{equation}
U(\tau) = e^{\kappa \tau (a^\dag \sigma_- - a \sigma_+)} = \sum_{n \geq 0} \frac{(\kappa \tau)^n}{n!}(a^\dag \sigma_- - a \sigma_+)^n
\end{equation}
can be written as 
\begin{align}
\nonumber
&& U(\tau) = \cos(\kappa \tau \sqrt{a^\dag a}) |g\rangle \langle g | - a \sin(\kappa \tau \sqrt{a^\dag a})/\sqrt{a^\dag a} |e\rangle\langle g| \\
\nonumber
&&\quad + \cos(\kappa \tau \sqrt{a a^\dag}) |e\rangle \langle e| + a^\dag \sin(\kappa \tau \sqrt{a a^\dag})/\sqrt{a a^\dag} |g\rangle\langle e|
\end{align}
by noting that 
\begin{equation}
\nonumber
(a^\dag \sigma_- - a \sigma_+)^{2k} = (-1)^k \left(\left(a^\dag a \right)^k |g\rangle\langle g| + \left(a a^\dag \right)^k |e\rangle\langle e|\right),
\end{equation}
and
\begin{equation}
\nonumber
(a^\dag \sigma_- - a \sigma_+)^{2k+1} = (-1)^k \left(a^\dag \left(a a^\dag \right)^k \sigma_- - a \left(a^\dag a \right)^k \sigma_+\right).
\end{equation}
Hence, the following initial states read
\begin{align}
\nonumber
U(\tau) |g,0\rangle & \rightarrow |g,0\rangle,\\
\nonumber
U(\tau)  |g,1\rangle& \rightarrow \cos(\kappa \tau)  |g,1\rangle -  \sin(\kappa \tau) |e,0\rangle,\\
\nonumber
U(\tau)  |e,0\rangle &\rightarrow \cos(\kappa \tau)  |e,0\rangle + \sin(\kappa \tau) |g,1\rangle.
\end{align}

\subsubsection{Absorption}
Let us now consider a collection of $N$ atoms, all prepared in the ground state $|g\rangle$ and each interacting with a single photon through the Jaynes-Cummings interaction. The state of the atoms associated with a successful absorption is given by 
\begin{equation}
\nonumber
\rho_{\text{cond}} = \tr_{\text{light}} \left[\ket{0}\bra{0} \otimes \id \-\ U_N \hdots U_1 |g\hdots g,1\rangle \langle g\hdots g,1 | U_1^\dag \hdots U_N^\dag\right] 
\end{equation}
and takes the form $\rho_{\text{cond}} = |\Psi_{\text{cond}}\rangle\langle \Psi_{\text{cond}}|$ when applying explicitly the $N$ propagators, where
\begin{equation}
\nonumber
|\Psi_{\text{cond}}\rangle =c^{N-1}s |g\hdots g e\rangle + c^{N-2} s |g\hdots e g \rangle + \hdots + s |e g \hdots g\rangle.
\end{equation}
Note that we have introduced the shorthands $c=\cos(\kappa \tau)$ and $s=\sin(\kappa \tau).$ The normalization of $\Psi_{\text{cond}},$ that is $1-\cos^{2N}(\kappa \tau),$  gives the probability of a successful absorption. For a small absorption amplitude per atom $\kappa \tau = \sqrt{d /N} \ll 1$ where $d=\alpha L$ is the total optical depth of the atomic chain and a large atom number, we have 
\begin{equation}
\lim_{N\rightarrow \infty} \cos^{2N}(\kappa \tau) \approx \lim_{N\rightarrow \infty} \left(1-\frac{d}{2N}\right)^{2N} \rightarrow e^{-d}
\end{equation}
which corresponds to the Bouguer-Beer-Lambert absorption law (eq.\ref{bouguer}). Similarly, the absorption probability $1- \cos^{2N}(\kappa \tau)$ tends to $1-e^{-d}.$

\subsubsection{Storage and retrieval probability}
The overall efficiency including the storage and retrieval probabilities is obtained by calculating 
\begin{equation}
\nonumber
|\langle g\hdots g, 1| U_N \hdots U_1 |\Psi_{\text{cond}}\rangle|^2.
\end{equation}
Note that we here consider a forward emission in which the retrieved photon is emitted in the same direction that the input photon. We obtain
\begin{align}
\nonumber
&&|\langle g\hdots g, 1| U_N \hdots U_1 |\Psi_{\text{cond}}\rangle|^2 = N^2 s^4 c^{2N-2} \\
&&  \approx d^2 \left(1-\frac{d}{N}\right)^{2N-2} \rightarrow d^2 e^{-d} \-\ \text{when} \-\ N \rightarrow \infty
\end{align} 
and thus retrieve the semi-classical forward efficiency eq.\eqref{eta_crib}.

For a backward emission, we obtain
\begin{equation}
|\langle g\hdots g, 1| U_1 \hdots U_N |\Psi_{\text{cond}}\rangle|^2 \rightarrow (1-e^{-d})^2 \-\ \text{when} \-\ N \rightarrow \infty
\end{equation}
corresponding to the semi-classical backward efficiency eq.\eqref{eta_crib_back}.

\subsubsection{Amplification through an inverted atomic ensemble} This simple model allows {us} to compute the expected noise of protocols for which the excited states are significantly populated when the stored excitation is released as in the two-pulse photon echo (2PE) protocol described in section \ref{2PE}. Consider first the case where all the atoms are in $|e\rangle$ and the field is in the vacuum state $|0\rangle.$ The mean photon number after an interaction time $\tau$ is given by 
\begin{equation} 
\nonumber
\langle e \hdots e,0 | U_N^\dag(\tau) \hdots U_1^\dag(\tau)  a^\dag a U_1(\tau) \hdots U_N(\tau) | e \hdots e,0\rangle
\end{equation}
which can be seen as the square of the norm of $a U_1(\tau) \hdots U_N(\tau) | e \hdots e,0\rangle.$ In the regime where the population remains essentially in the excited state, the atomic operators $\sigma_\pm$ verifies
\begin{equation}
[\sigma_+, \sigma_-] = |e\rangle\langle e| - |g\rangle \langle g| \approx 1.
\end{equation}
In this case, the Hamiltonian \eqref{Jaynes_Cummings} is a squeezing operator between two bosonic modes and the formula $e^B A e^{-B} = \sum_{n\geq 0} \frac{1}{n!} \underbrace{[B,\hdots [B, A]\hdots ]}_{\text{n times}}$ can be used to prove that
\begin{equation}
a U_1 = U_1 \left(\cosh(\kappa \tau) a + \sinh(\kappa \tau) \sigma_-^{(1)}\right).
\end{equation}
Commuting $a$ with $U_2 \hdots U_N,$ we obtain
\begin{align}
\label{transform_field_s}
&a U_1  \hdots  U_N = U_1  \hdots  U_N \Big(\cosh(\kappa \tau)^N a \\
\nonumber
&+ \cosh(\kappa \tau)^{N-1} \sinh(\kappa \tau) \sigma_-^{(N)} + \hdots +\sinh(\kappa \tau) \sigma_-^{(1)} \Big)
\end{align}
where $\sigma_-^{(i)}$ is the atomic operator $\sigma_-$ for the $i^{th}$ atom. This leads to  
\begin{align}
\nonumber
& || a U_1(\tau) \hdots U_N(\tau) | e \hdots e,0\rangle ||^2 \\
\nonumber
&= \sinh(\kappa \tau)^2 \sum_{j=1}^N \cosh(\kappa \tau)^{2j-2} = \cosh(\kappa \tau)^{2N}-1.
\end{align}
Using $\kappa \tau = \sqrt{d/N} \ll 1$ and taking the limit of large $N$, the mean photon number is
\begin{equation}
e^d-1.
\end{equation}
This corresponds to the number of photons emitted in a single mode by an inverted ensemble \cite{RASE, Sekatski}.

More generally, eq.\eqref{transform_field_s} shows that in the regime where the atoms are mainly in the excited state, the atomic ensemble operates as a classical amplifier, the gain $G$ depending exponentially on the optical depth via $G=e^{d}.$ Such an amplifier transforms the field operators according to
\begin{align}
\label{a_out}
& \bar U^\dag a \bar U = \sqrt{G} a + \sqrt{G-1} \-\ \sigma_c^\dag \\
\label{adag_out}
& \bar U^\dag a^\dag \bar U = \sqrt{G} a^\dag + \sqrt{G-1} \-\ \sigma_c 
\end{align}
where $\bar U^\dag a \bar U  = U_N^\dag (\tau) \hdots U_1^\dag (\tau) a U_1(\tau) \hdots U_N(\tau).$ 
The bosonic operators $\sigma_c$ and $\sigma_c^\dag$ annihilates and creates collectively atoms in the ground state
$$
\sigma_c^\dag = \frac{1}{\sqrt{M}}(\cosh(\kappa \tau)^{N-1} \sinh(\kappa \tau) \sigma_-^{(N)} + \hdots +\sinh(\kappa \tau) \sigma_-^{(1)})
$$
 with the normalization coefficient $M=\cosh(\kappa \tau)^{2N}-1.$\\
 
Equations \eqref{a_out}\&\eqref{adag_out} allow one to derive expectations values for the field when the atoms are mostly excited. Let us consider for example a 2PE where the first pulse is a single photon Fock state and the second pulse is a $\pi$-pulse. We can compute the response of a non-photon number resolving detector with dectection efficiency $\eta$ at the echo time. If the photon has been successfully absorbed, at the echo time, the atoms are well described by a single excitation $|1\rangle$ in the collective mode $\sigma_c$ and the field $a$ is in the vacuum state. The probability that the photon detector clicks is given by 
\begin{align}
\nonumber
\langle 01| \bar U^\dag \big(1-(1-\eta)^{a^\dag a}\big) \bar U |01\rangle & = 1 - \langle 01| \bar U^\dag :e^{-\eta a^\dag a} : \bar U |01\rangle \\
\nonumber
& = 1- \langle 01| \bar U^\dag (1 -\eta aa^\dag + \frac{\eta^2}{2} a^\dag a ^\dag a a - \hdots) \bar U |01\rangle.
\end{align}  
Using  eqs.\eqref{a_out}\&\eqref{adag_out}, we easily show that 
$$ \langle \bar U^\dag \underbrace{a^\dag  \hdots a ^\dag}_{\text{k times}} \underbrace{a \hdots a}_{\text{k times}} \bar U\rangle = (k+1)! (G-1)^k.$$
Therefore 
\begin{align}
\nonumber
\langle 01| \bar U^\dag \big(1-(1-\eta)^{a^\dag a} \big) \bar U |01\rangle & = 1- \sum_{k\geq 0} (-1)^k \eta^k (G-1)^k (k+1)\\
\label{clic_2pulse}
& = 1- \frac{1}{\big(1+\eta(G-1)\big)^2}.
\end{align}
This formula shows that no click is obtained when the detection efficiency is null while the detectors clicks with unit probability as long as $\eta G \gg 1.$
 
\subsubsection{Beamsplitter interaction in a non-inverted ensemble}
In the regime where the atoms are and remain essentially in the ground state, the atomic operators $\sigma_\pm$ verifies
\begin{equation}
[\sigma_+, \sigma_-] = |e\rangle\langle e| - |g\rangle \langle g| \approx -1.
\end{equation}
As in the previous paragraph, the formula $e^B A e^{-B} = \sum_{n\geq 0} \frac{1}{n!} [B, A]^n$ can thus be used to prove that in this regime
\begin{equation}
a U_N = U_N \left(\cos(\kappa \tau) a + \sin(\kappa \tau) \sigma_-^{(N)}\right).
\end{equation}
We thus have  
\begin{align}
\label{transform_field}
&a U_N  \hdots  U_1 = U_N  \hdots  U_1 \Big(\cos(\kappa \tau)^N a \\
\nonumber
&+ \cos(\kappa \tau)^{N-1} \sin(\kappa \tau) \sigma_-^{(1)} + \hdots + \sin(\kappa \tau) \sigma_-^{(N)} \Big).
\end{align}
By introducing the collective operator 
\begin{equation}
\bar{\sigma}_c = \frac{1}{\sqrt{\overline M}}(\cos(\kappa \tau)^{N-1} \sin(\kappa \tau) \sigma_-^{(N)} + \hdots +\sin(\kappa \tau) \sigma_-^{(1)})
\end{equation}
with $\overline M = 1-\cos(\kappa \tau)^{2N},$ the atom-light interaction can be seen as a standard beamsplitter-type interaction
\begin{align}
\label{a2}
&  \bar U^\dag a \bar U = \sqrt{e^{-d}} a + \sqrt{1-e^{-d}} \-\ \bar{\sigma}_c \\
\label{adag2}
& \bar U^\dag a^\dag \bar U = \sqrt{e^{-d}} a^\dag + \sqrt{1-e^{-d}} \-\ \bar{\sigma}_c^\dag.
\end{align}

The formulas derived from this simple quantum model will be helpful to characterize the quantum nature of different storage protocols as we will see now.
\subsection{Continuous variable criterion}\label{CV_criterion}
Here, we study the propagation and read-out of a pulse with quantum noise through different memories and {review} a criterion to evaluate if {the output state is the best cloned copy of the input, that is, to guarantee that no better copy of the input state is available}. We analyze generic storage protocols in a continuous variable perspective to estimate the amount of noise and loss that can be tolerated {to fulfill this criterion}.

\subsubsection{The stored quantum states}

A quantum memory should be able to store and retrieve any state {while preserving its quantum features}. The state can be a classical state but its quantum statistics should be preserved. In continuous variable quantum information, the variables of interest are the field quadratures, defined as
\begin{equation}X^+=a+a^\dagger\end{equation} and 
\begin{equation}X^-=-i(a-a^\dagger)\end{equation}
where $a$ and $a^\dagger$ are the creation and annihilation operators of the field, as in the previous section. As they satisfy the canonical commutation relations $[a,a^\dagger]=1$, it follows that $[X^+,X^-]=2i$ and that
 \begin{equation}n=\frac{1}{4}\big[(X^+)^2+(X^-)^2\big]-\frac{1}{2}\end{equation}
where $n=a^\dagger a$ is the photon number operator.

The signal at the output of a quantum memory can be decomposed into a classical amplitude $\alpha$ and a fluctuating noise term $\delta\hat{X}^\pm$.
Formally, for a gaussian state, we write the amplitude and phase quadratures of the field as
\begin{equation}\hat{X}^\pm=\alpha^\pm+\delta\hat{X}^\pm \end{equation}

To avoid writing the propagators when describing the field at the output of the memory, we now introduce the subscript $_{\rm {out}}$ defined as  $a_{\rm {out}}= \overline{U}^\dagger a\overline{U}$ for example. Similarly, the subscript $_{\rm {in}}$ is used to describe the input of the memory.
The measured output signal is generally the power spectral density, given by the Fourier transform of the autocorrelation function.
It reads as
\begin{equation}S_{\rm out}^\pm=\langle (\hat{X}_{\rm out}^\pm)^2 \rangle \end{equation}
and the noise  as
\begin{equation}V_{\rm out}^\pm=\langle \delta (\hat{X}_{\rm out}^\pm)^2 \rangle \end{equation}

We thus obtain
\begin{equation}S_{\rm out}^\pm=(\alpha_{\rm out}^\pm)^2+V_{\rm out}^\pm.\end{equation}
We will estimate $\alpha_{\rm out}^\pm$ and $V_{\rm out}^\pm$ at the output of the optical memories and identify the values that enable entering the quantum memory regime.

\subsubsection{Quantum memory criterion}

Generally, optical memories are benchmarked against quantum information criteria. In particular, the performance of a given quantum memory can be {evaluated similarly to a quantum teleportation scheme} {by quantifying the quality of the output state with respect to the input.}

\begin{figure}[ht!]
\centerline{\scalebox{0.3}{\includegraphics{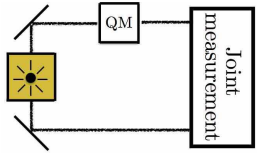}}}
\caption{General scheme for characterizing an optical memory. A pair of EPR entangled beams with a mean signal amplitude is prepared. One of these beams is injected into, stored, and readout from the optical quantum memory (QM) while the other is being propagated in free space. A joint measurement with appropriate delay is then used to measure the quantum correlations between the quadratures of the two beams. 
}\label{carct}
\end{figure}

Figure \ref{carct} shows the schematics of the quantum memory benchmark. The optimal classical measure and prepare strategy for optical memory consists in measuring the input state jointly on two conjugate quadratures using two homodyne schemes \cite{Hammerer}. The measured information is stored before fed-forward onto an independent beam. In this classical scheme, the storage time can be arbitrarily long without additional degradation. However, two conjugate observables cannot be simultaneously measured and stored without paying a quantum of duty. Moreover, the encoding of information onto an independent beam will also introduce another quantum of noise. In total, the entire process will incur two additional quanta of noise onto the output optical state \cite{HetetPRA}.

Characterizing quantum memory using the state-dependent fidelity as a measure can be complicated for exotic mixed states. Alternatively, we use the signal transfer coefficients $T$ and the input-output conditional variances $V_{cv}$ to establish the efficiency of a process \cite{Grangier, Ralph}. The conditional variances and signal transfer coefficients are defined as 

\begin{equation} V_{cv}^{\pm}=V_{\rm out}^{\pm} -\frac{ |\langle X^\pm_{\rm in} X^\pm_{\rm out} \rangle|^2}{V^\pm_{\rm in}} \end{equation}

and 

\begin{equation} T^\pm = \frac{R^\pm_{\rm out}}{R^\pm_{\rm in}} \end{equation}

where $R^\pm_{\rm out/in}$ is defined as

\begin{equation} R^\pm_{\rm out/in}=\frac{4(\alpha^\pm_{\rm out/in})^2}{V^\pm_{\rm out/in}}. \end{equation}

We now define two parameters that take into account the performances of the system on both conjugate observables as 

\begin{equation} V=\sqrt{V_{cv}^{+}V_{cv}^{-}} \end{equation}
and \begin{equation} T=T^++T^- \end{equation}

It can be shown that a classical memory {based on the measure and prepare scheme described before} cannot overcome the $T>1$ or $V<1$ limits \cite{Ralph}. With a pair of entangled beams, it is possible to have an output state with $V<1$ or $T>1$, hence demonstrating that the memory outperforms the optimal measure and prepare strategy. In case where the output state satisfies both $V<1$ and $T>1,$ the output is the best possible cloned copy of the input state \cite{Grosshans}.
A perfect quantum memory would satisfy both $T=2$ and $V=0$.

\subsubsection{{\it Slow-light} memory}\label{CV_SL}

\begin{figure}[hb!]
\centerline{\scalebox{0.9}{\includegraphics{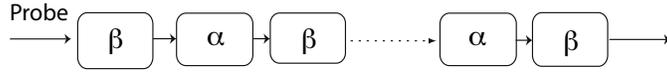}}}
\caption{Field propagating in a medium with gain $\alpha$ and loss $\beta$.
}\label{Setup}
\end{figure}

We now present a general theory for amplification and attenuation of a traveling wave and use the TV diagram to quantify the amount of excess noise that is tolerated.
This theory is well adapted to {\it slow-light} memories \cite{HetetPRA} but can be carried over to other memories, like Raman (section \ref{Raman}) or CRIB (section \ref{CRIB}) memories. Gain can indeed be present if, for instance, population has been transferred to other states during the mapping and read-out stages.

As discussed in the previous section, and in particular in eqs.\eqref{a_out}\&\eqref{adag_out}, the output of an ideal linear amplifier with a gain factor $G>1$, relates to the input field {\it via} this relation :
\begin{equation} a_{\rm out}=\sqrt{G} a_{\rm in}+\sqrt{G-1}\sigma_c^\dagger \end{equation}
where $\sigma_c^{\dagger}$ is a bosonic operator in the vacuum state. The power
spectrum at the output of an ideal phase-insensitive amplifier
is then given by \begin{equation}S_{\rm out}=GS_{\rm in} +G-1 \end{equation} where $S_{\rm in}$ is input spectrum.

By concatenating $m$ amplifying and attenuating infinitesimal slices with linear amplification $1+\alpha \delta z$ and attenuation
$1-\beta \delta z$ where $\delta z=z/m$, as represented in fig.\ref{Setup}, we will calculate the noise properties of the field. 

\begin{figure}[ht!]
\centerline{\scalebox{0.7}{\includegraphics{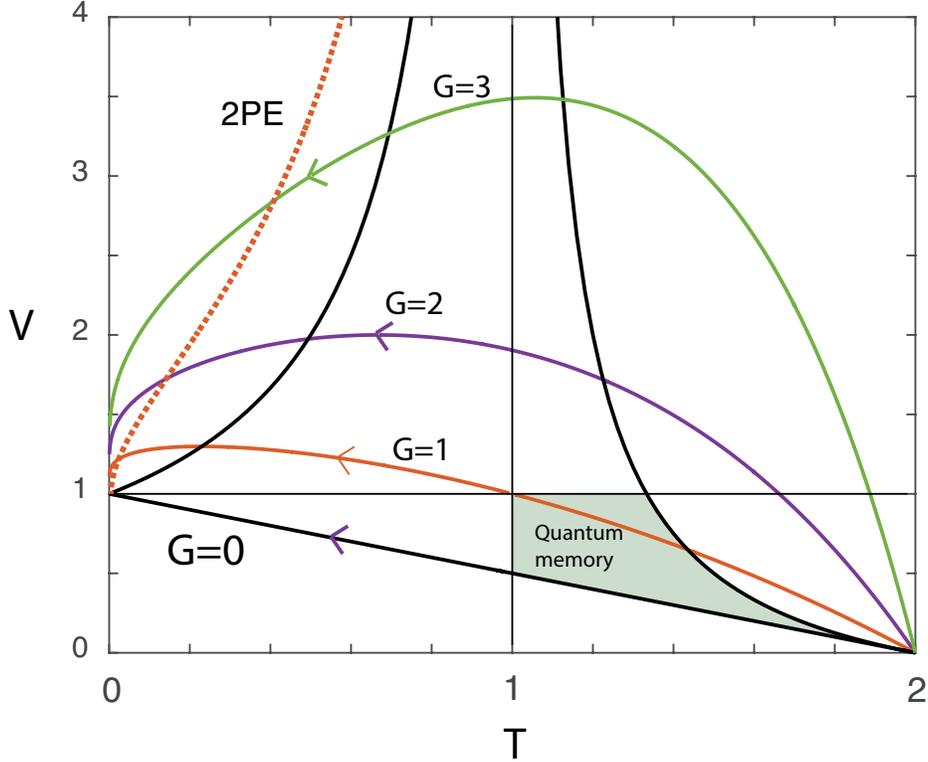}}}
\caption{TV diagram for a CRIB memory with varying gain and loss. The dashed line shows the evolution of the standard 2PE memory performance as a function of optical depth.
}\label{TV}
\end{figure}

The power spectrum of the field at a slice $m$ is
\begin{equation}
S_m=(1+\frac{(\alpha-\beta)z}{m})^m (S_{in}-1) +1+2\alpha\sum (1+\frac{(\alpha-\beta)z}{m})^{m-j}
\end{equation}
\\
In the infinitesimal slice width limit, we obtain
\begin{equation} S_{\rm out}=\eta S_{\rm in} +(1-\eta)(1+N_f) \end{equation}
where $N_f=2\alpha/(\beta-\alpha)$ and $\eta=\exp{((-\beta+\alpha)L)}$ where $L$ is the length of the medium.
Using standard memory protocols, one can find a relationship between $\alpha$, $\beta$ and the memory parameters.  
One can then show that 
\begin{equation} V=1-\eta+V_{\rm noise} \end{equation}
and \begin{equation} T=2\eta/(1+V_{\rm noise}),\end{equation} where $V_{\rm noise}=1+(1-\eta) N_f$.
Figure \ref{TV} shows a TV diagram for a memory with varying loss (arrows) and three different gain values.

If there is no mean intensity of the field at the input, the output field is simply the memory output noise. It reads as 
\begin{equation} S_{\rm out}=\eta +(1-\eta) (1+N_f) \end{equation}
If we further assume that all the atoms are in the excited state, that is the atomic
medium operates as an amplifier ($\beta$=0), $N_f=-2$, now we obtain  
\begin{equation} S_{\rm out}=1-2(1-\eta)=2\eta-1 \end{equation}

Assuming that the noise is the same on both quadratures, and the relation between the mean number of photons and the field quadratures leads to 
 \begin{equation} \langle n \rangle=\frac{1}{2}\big[\langle X ^2 \rangle-1\big]
 = \frac{1}{2}\big[S_{\rm out}-1\big]=\eta-1
\end{equation}

The mean number of photon is thus
\begin{equation} \label{mean_photon_number}
e^{d}-1
\end{equation}
where $d=\alpha L$ in the optical depth, as was found in the previous section.
\begin{figure}[ht!]
\centerline{\scalebox{0.55}{\includegraphics{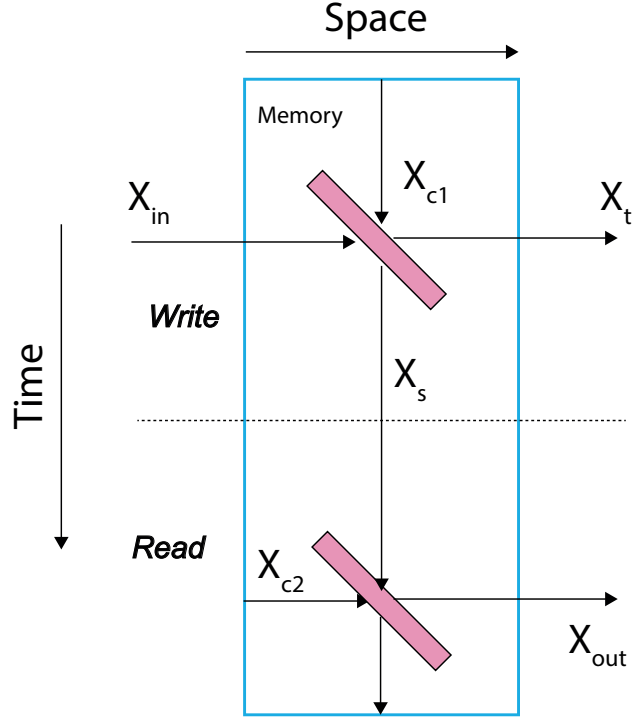}}} 
\caption{Beam splitter description of photon echo memories.}\label{BS}
\end{figure}

\subsubsection{Photon echo memories}

Standard photon echo protocols that use long lived excited state transitions are generally not immune to noise. If the emission takes place while population remains in the excited state, gain will be present so the memory will not enter the quantum regime.

\paragraph{Controlled reversible inhomogeneous broadening}

The CRIB scheme can be modeled using arrays of beam-splitters. In its most efficient form, namely using the gradient echo memory scheme (GEM) \cite{hetet2008electro} (sometimes called longitudinal CRIB) or using (transverse-) CRIB with a backward write pulse 
(section \ref{CRIB}), the write and read stages can be seen as two beam-splitters with a reflectivity that depends on the optical depth \cite{LongdellAnalytic}, as depicted in fig.\ref{BS}.
Let us note that without a backward pulse, more beam-splitters are needed to describe the output field of the CRIB memory \cite{LongdellAnalytic}. 
In these scheme, the population remains mainly in the ground state so that gain, and thus noise, will be absent.

For the write stage, the transmitted pulse field intensity is attenuated according to the Bouguer-Beer-Lambert law by $e^{-d}$.
In terms of quadratures, including the "vacuum port" modeling atomic fluctuations, we deduce from eqs.\eqref{a2}\&\eqref{adag2} the expressions for the light and spin quadratures at the two output ports as defined in fig.\ref{BS}

\begin{equation} X_t=\sqrt{e^{-d}} X_{\rm in} - \sqrt{1-e^{-d}}X_{c1} \end{equation}
and
\begin{equation} X_s=\sqrt{1-e^{-d}} X_{\rm in} + \sqrt{e^{-d}}X_{c1} \end{equation}

The vacuum contribution ensures preservation of the commutation relations of the field and atomic operators. 

In the case of CRIB with backward propagation or GEM (forward), the beam-splitter reflectivity is "inverted" and the output field can be written simply as 

\begin{equation} X_{\rm out}=\sqrt{1-e^{-d}} X_{s} + \sqrt{e^{-d}}X_{c2}.\end{equation}
In the absence of signal, the input is in the vacuum state $V_{\rm in}=1$,  so that $V_{\rm out}=1$.
We thus find a conditional variance
\begin{equation} V_{cv}^{\pm}=1-(1-e^{-d})^2 \end{equation}
and transfer coefficient.
\begin{equation} T=2 (1-e^{-d})^2.\end{equation}

So for the present case of CRIB and in the limit of large optical depth, $T\rightarrow 2$ and $V\rightarrow 0$, the CRIB memory is a quantum memory, as represented by the green area in fig.\ref{TV}.

\paragraph{Standard two-pulse photon echo}

Let us now consider the 2PE memory described in section \ref{2PE}. The difference between the 2PE memory and the CRIB is that the atoms are in the excited state during the read-out retrieval stage. This inversion implies that the input light will be amplified, which will invariably add noise.

Considering again the two-beam-splitter approach depicted in fig.\ref{BS}, the writing stage is the same as CRIB, so a fraction $\sqrt{\eta_R}=\sqrt{1-e^{-d}}$ of the field is written in the memory.
The output field however is amplified by a quantity $\sqrt{\eta_W}=\sqrt{e^d-1}$ as discussed in the {\it slow-light} section \ref{CV_SL}.
In total, the transmission thus reads \begin{equation}\eta_R \eta_W=(1-e^{-d})(e^d-1)=4~{\rm sinh}^2(d/2)\end{equation}We here retrieve the semi-classical 2PE efficiency eq.\eqref{etaPi}.

In terms of the field quadratures, we have
\begin{equation} X_{\rm s}=\sqrt{1-e^{-d}} X_{\rm in}+\sqrt{e^{-d}}X_{\rm c1}\end{equation}
and 
\begin{equation} X_{\rm out}=\sqrt{e^{d}-1} X_{\rm s}+\sqrt{e^{d}}X_{\rm c2}\end{equation}
just like for a linear amplifier (eqs.\ref{a2}\&\ref{adag2}).

The product of the conditional variances is thus 
\begin{equation}V=1-e^{-d}+e^d\end{equation}
and the sum of the two signal-to-noise transfer coefficients is 
\begin{equation}T=\frac{4 {\rm sinh}(d/2)^2}{2 e^d-1}\end{equation}

These two quantities are plotted in fig.\ref{TV} (dotted line), showing the 2PE memory does not enter the quantum regime for any optical depth.
We have checked that a better performance (a lower V and larger T) can be obtained if the optical depth is lowered during the writing stage but the memory would still operate in the classical domain. 

\subsection{Photon counting criteria}\label{counting_criterion}
We now present various criteria for certifying the quantum nature of storage protocols based on photon counting, including the autocorrelation measurement, the Cauchy-Schwarz criterion and the Bell test.

\subsubsection{Autocorrelation measurement}
Let us consider a single-mode of the electromagnetic field with bosonic operators $a$ and $a^\dag$ and described by the state $\rho_a.$ This state is said classical if it can be represented as a mixture of coherent states $|\alpha\rangle$, that is, one can find a quasi-probability distribution $P(\alpha) \geq 0$ such that 
\begin{equation}
\rho_a=\int d^2 \alpha P(\alpha) |\alpha\rangle\langle \alpha|.
\end{equation}
The autocorrelation of this field defined as
\begin{equation}
g_a^{(2)}=\frac{\langle a^{\dag 2}  a^2 \rangle}{\langle a^\dag a \rangle^2}
\end{equation}
is at least equal to 1 \cite{loudon2000quantum}. Conversely, if the result of an autocorrelation measurement is smaller than 1, one can conclude that the measured state is non-classical. A single photon Fock state for example, is a non-classical state because its autocorrelation is 0. A simple way to certify the quantum nature of a given memory is thus to store a single photon and to check that the result of an autocorrelation measurement after retrieval is smaller than 1. This shows that the memory preserves the non-classical feature of light.

\begin{figure}
\centering
\includegraphics[width=0.5\textwidth]{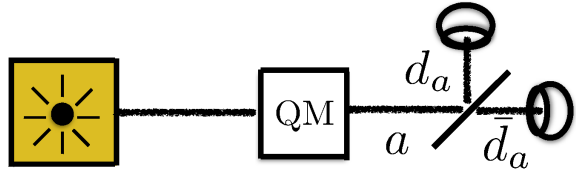}
\caption{Certifying the quantum nature of a memory by checking that it can preserve the non-classical property of a single-mode field. A single photon Fock state is stored and an autocorrelation measurement is performed on the retrieved photon.}
\label{Fig1}
\end{figure}

Note that in practice, non photon-number resolving detectors can be used to certify the non-classical nature of a single-mode field: it is sufficient to put two of these detectors after a 50/50 beamsplitter and to check that the probability of a twofold coincidence is smaller than the product of probabilities of singles \cite{sekatski2012detector}. Let us thus consider the experiment represented in fig.\ref{Fig1} where a source produces a single photon that is subsequently stored in a memory. The photon is then released and an autocorrelation measured with non photon-number resolving detectors ($d_a$ and $\bar{d}_a$) with efficiency $\eta_d$ each. Let $\eta_m$ be the efficiency of the memory and $p_{{\text{dc}}}$ the probability to get a dark count, that is a click on one detector when the photon source is switched off. Obviously, $\eta_m$ can include the non-unit efficiency of the source and the loss from the source to the memory. $\eta_d$ also accounts for the loss between the 50/50 beamsplitter and each detector. $p_{{\text{dc}}}$ includes the detector dark counts and various sources of noise operating independently on each detector. We assume that $p_{{\text{dc}}}$ is the same for both detectors ($d_a$ and $\bar{d}_a$). To obtain twofold coincidences smaller than the product of singles, these parameters has to fulfill the following inequality  (see appendix \ref{appendix:formulas_counts} for details)
\begin{equation}
\label{autocorrelation}
g_a^{(2)}=\frac{1-2(1-p_{\text{dc}})(1-\eta_d\eta_m/2)+(1-p_{\text{dc}})^2(1-\eta_d\eta_m)}{\left(1-(1-p_{\text{dc}})(1-\eta_d\eta_m/2)\right)^2} < 1.
\end{equation} 
Note that in the absence of noise ($p_{\text{dc}}=0$), this ratio is zero independently of the efficiency. In other words, for an ideal implementation of a memory protocol without noise, there is no constraint on the memory efficiency to prove that it can preserve the result of an autocorrelation measurement performed on a single photon. For unit efficiencies $\eta_d\eta_m =1,$ the ratio eq.\eqref{autocorrelation} tends to $1-\epsilon^2/4$ for $p_{\text{dc}} \approx 1-\epsilon.$ For low efficiencies $\eta_d\eta_m \ll 1,$ the inequality \eqref{autocorrelation} is fulfilled as long as  ${p_\text{dc}} \leq 3\eta_d\eta_m.$ 

It is worth mentioning that the here proposed criterion does not allow to conclude that the memory outperforms classical strategies for storage as a device that would throw the photon emitted by the source and create a photon afterward would lead to a zero autocorrelation measurement. However, under the assumption that memory under test indeed operates as a storage/retrieval protocol, this criterion shows that this memory is in the quantum regime, in the sense that it preserves the non-classical nature of a single mode field.

It is interesting to compute the result of an autocorrelation measurement that would be obtained by storing a single photon in an atomic ensemble using the 2PE technique. In the ideal scenario where there is no loss before and inside the memory and the photon absorption is successful, we find for the autocorrelation
\begin{equation}
\label{g22pe}
\frac{1-\frac{2}{(1+\frac{\eta_d}{2}(e^d-1))^2} + \frac{1}{(1+\eta_d(e^d-1))^2}}{\left(1-\frac{1}{(1+\frac{\eta_d}{2}(e^d-1))^2}\right)^2}
\end{equation}
which tends to $3/2$ for small optical depth $d\ll 1$ and to 1 when $\eta/2 (e^{d}-1) \geq 1.$ Even in an ideal scenario, we conclude that a storage technique based on a two-pulse photon echo does not preserve the non-classical nature of a single photon.

\subsubsection{Cauchy-Schwarz criterion}
The Cauchy-Schwarz parameter R can be used to reveal non-classical correlations between two fields \cite{clauser1974experimental}. Consider two single-mode fields and their respective bosonic operators $a,$ $a^\dag$ and $b,$ $b^\dag.$ We say that these fields are classically correlated if their state $\rho_{ab}$ can be written as a mixture of coherent states $|\alpha\rangle$, $|\beta\rangle,$ that is, there exists a non negative function $P(\alpha, \beta)$ such that 
\begin{equation}
\rho_{ab}=\int d^2\alpha d^2\beta P(\alpha, \beta) |\alpha, \beta\rangle\langle \alpha, \beta|. 
\end{equation}
The Cauchy-Schwarz parameter defined as 
\begin{equation}
R=\frac{\langle a^\dag b^\dag b a \rangle}{ \langle a^{\dag 2} a^2 \rangle \langle b^{\dag 2} b^2 \rangle}
\end{equation}
is at most equal to 1 when calculated on classically-correlated states. $R > 1$ is a witness of non-classical correlations.

\begin{figure}
\centering
\includegraphics[width=0.75\textwidth]{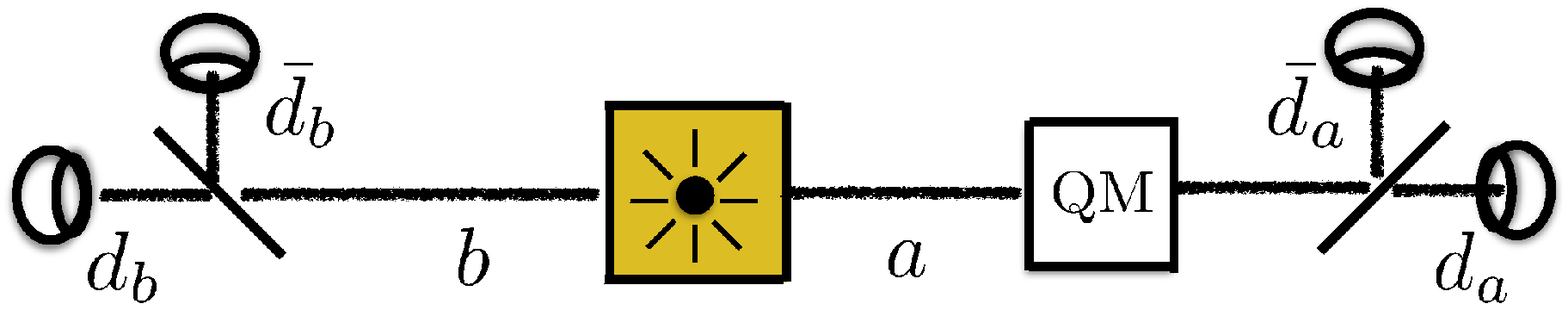}
\caption{Setup to certify the quantum nature of a memory by checking that it can preserve the non-classical correlations between two fields. A photon pair source is used to produce two-mode squeezed vacuum states. One of the two modes (mode $a$) is stored in a memory and the Cauchy-Schwarz parameter is measured between the mode $a$ after retrieval and the mode $b.$}
\label{Fig2}
\end{figure}

As for the autocorrelation measurement, the Cauchy-Schwarz parameter can be measured with non photon-number resolving detectors \cite{sekatski2012detector}, see fig.\ref{Fig2}. It is sufficient to take the ratio between the square of twofold coincidences between detectors $d_a$\&$d_b$ and the product of coincidences between $d_a$\&$\bar{d}_a$ and $d_b$\&$\bar{d}_b.$ Let us consider the experiment shown in fig.\ref{Fig2} with a source producing two-mode vacuum squeezed states, that is 
\begin{equation}
(1-p)^{\frac{1}{2}} e^{\sqrt{p} a^\dag b^\dag} |00\rangle.
\end{equation}
Further consider the storage and release of the mode $a$ into a memory with efficiency $\eta_m.$ Let $\eta_{d}^a$ ($\eta_{d}^b$) be the efficiency of detectors $d_a$ and $\bar{d}_a$ ($d_b$ and $\bar{d}_b$) and $p_{\text{dc}}^a$ ($p_{\text{dc}}^b$) the probability to get a click on the detector $d_a$ or $\bar d_a$ ($d_b$ or $\bar d_b$) when the source is switched off (dark counts). As before, the memory efficiency includes the loss from the source to the memory. The efficiency of detectors $d_a$ and $\bar{d}_a$ includes the loss from the beamsplitter to the detector while the efficiency of the detectors $d_b$ and $\bar{d}_b$ includes the loss from the source to the detector (without the transmission of the beamsplitter).  $p_{\text{dc}}^a$ which we assume to be the same for the two detector $d_a$ and $\bar{d}_a,$ includes various source of noise that can be modeled as detector dark counts. In this scenario, the Cauchy-Schwarz parameter is given by  (see appendix \ref{appendix:formulas_counts} for details)
\begin{align}
\label{Cauchy_Schwarz}
R=&\Bigg[1-\frac{(1-p_{\text{dc}}^a)(1-p)}{1-p(1-\frac{\eta_d^a\eta_m}{2})}-\frac{(1-p_{\text{dc}}^b)(1-p)}{1-p(1-\frac{\eta_d^b}{2})}
+\frac{(1-p_{\text{dc}}^a)(1-p_{\text{dc}}^b)(1-p)}{1-p(1-\frac{\eta_d^a\eta_m}{2})(1-\frac{\eta_d^b}{2})}\Bigg]^2/\nonumber \\
\nonumber
& \Bigg[\left(1-2\frac{\left(1-p_{\text{dc}}^{a}\right)(1-p)}{1-p(1-\frac{\eta_d^a\eta_m}{2})}+\frac{\left(1-p_{\text{dc}}^{a}\right)^2(1-p)}{1-p(1-\eta_d^a\eta_m)}\right)\times\\
&\left(1-2\frac{\left(1-p_{\text{dc}}^{b}\right)(1-p)}{1-p(1-\frac{\eta_d^b}{2})}+\frac{\left(1-p_{\text{dc}}^{b}\right)^2(1-p)}{1-p(1-\eta_d^b)}\right)\Bigg]
\end{align}
and has to be larger than $1$ to certify that the tested memory preserves non-classical correlations.  In the ideal case with unit efficiencies and no dark count, the Cauchy-Schwarz parameter tends to $\frac{1}{4}(1+\frac{1}{p})^2$ for $p \ll 1$. Note that $p$ can be written as a function of the mean photon-number emitted in one mode ($a$ or $b$) as $p=n/(n+1).$

The Cauchy-Schwarz criterion leads to similar conclusions than the autocorrelation measurement. If the memory under test is a device that throws the incoming field away and produces a single photon at a later time, the Cauchy-Schwarz parameter would tend to infinity, independently of the state of mode $b$. However, assuming that the tested memory indeed operates as a storage/retrieval protocol, the Cauchy-Schwarz criterion allows to conclude that the memory preserves non-classical correlations between two fields.

\subsubsection{Bell test}

The two criteria presented previously do not test the capability of a memory to preserve the coherence. This can be done by storing a part of an entangled state and by checking that the entanglement is preserved using, for example, a Bell test. A possible realization would use a photon pair source emitting entangled photon pairs, for example, in polarization. The spatial mode $a$ is stored in a memory and subsequently released. Measurements are finally performed combining wave-plates, polarizing beamsplitters and one detector on each side. The twofold coincidences are recorded. Two interference patterns are obtained, one by rotating the analyzer on the left side, the other one by rotating the analyzer on the right side. If the mean visibility of this interference patterns is larger than $1/3,$ one can conclude about the presence of entanglement under the assumption that the state is a mixture between the singlet state and white noise. As the memory operates as a local operation, which cannot increase entanglement, a high interference visibility witnesses the presence of entanglement between the photon in $b$ and the excitation stored in the memory. Note that there is no need to close the detection and locality loopholes here as the Bell test is used as an entanglement witness, not as non-locality Bell test. 

\begin{figure}
\centering
\includegraphics[width=0.75\textwidth]{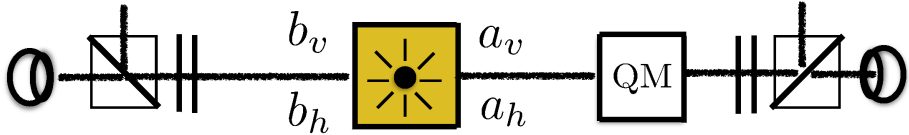}
\caption{Setup to certify the quantum nature of a memory by checking that it can preserve entanglement between two fields. A photon pair source is used to produce entangled photon pairs in polarization. Two of the four modes (mode $a_h$ and $a_v$) are stored in a memory and a Bell inequality violation can be inferred from the visibility of the interference that is obtained by recording the twofold coincidences while rotating the measurement settings locally.}
\label{Fig3}
\end{figure}

Let us consider the experimental realization shown in fig.\ref{Fig3} with a source based on spontaneous parametric down conversion, that is, photon pairs described by
\begin{equation}
|\psi^-_{a_h a_v b_h b_v}\rangle = (1-p) e^{\sqrt{p} (a_h^\dag b_v^\dag-a_v^\dag b_h^\dag)} |00\rangle.
\end{equation}
Let $\eta_a$ and $\eta_b$ be the detector efficiency on side $a$ and $b$ respectively and $p_{\text{dc}}^a$ and $p_{\text{dc}}^b$ the corresponding noise. As before, the memory efficiency is labeled $\eta_m.$ The visibility of the interference $V$ is given by (see appendix \ref{appendix:formulas_counts} for details)
\begin{align}
\label{Bell}
V=\Bigg[\frac{(1-p)(1-p_{\text{dc}}^a)(1-p_{\text{dc}}^b)}{1-p(1-\eta_a)(1-\eta_b)} - 
 \frac{(1-p)^2(1-p_{\text{dc}}^a)(1-p_{\text{dc}}^b)}{(1-p(1-\eta_a))(1-p(1-\eta_b))}\Bigg]/\nonumber \\
 \Bigg[ 2-2\frac{(1-p_{\text{dc}}^a) (1-p)}{1-p(1-\eta_a)}-2\frac{(1-p_{\text{dc}}^b) (1-p)}{1-p(1-\eta_b)}+
 \frac{(1-p)(1-p_{\text{dc}}^a)(1-p_{\text{dc}}^b)}{1-p(1-\eta_a)(1-\eta_b)} +\nonumber \\
 \frac{(1-p)^2(1-p_{\text{dc}}^a)(1-p_{\text{dc}}^b)}{(1-p(1-\eta_a))(1-p(1-\eta_b))}
\Bigg].
\end{align}
As before, $p$ can be written as a function of the mean photon number in one mode ($a_h,$ $a_v,$ $b_h$ or $b_v$) as $p=n/(n+1).$

Interestingly, one can conclude from such a Bell test that the memory performs better that any possible classical strategies using for example a measure and prepare strategy or cloning followed by measurements in different basis. In this case, entanglement would be broken and the visibility would be limited to 1/3 assuming that the classical strategies introduce white noise on the singlet state.

Note that it has been shown recently that a device-independent certification is possible in the setup presented in fig.\ref{Fig3} \cite{sekatski_inprep}. In other words, it is possible to certify that the memory is a unitary operation and applies the identity on the qubits independently of the details and imperfections of the actual implementation by performing Bell tests with and without storage.
\\

We have derived and analyzed complementary criteria, both for continuous and discrete variables. They can be used as a benchmark to certify the quantum nature of the memory outcome. Our goal was to relate quantum optics measurements to experimentally accessible quantities that can be evaluated independently. This explicit criteria can be alternatively considered as a guide to anticipate the result of quantum measurements, to identify the limitations of an experimental setup and/or as an analytical tool for modeling a posteriori.

\section{Conclusion}

We have reviewed a series of quantum optical memory protocols conceived to store information in atomic ensembles. Without providing an exhaustive review of the different systems and techniques, we propose to put the protocols into two categories, namely photon echo and {\it slow-light} memories.
Our analysis is based on the significant differences of storage and retrieval dynamics. We have used a minimalist semi-classical Schr\"odinger-Maxwell model to describe the signal propagation and to evaluate the storage efficiency in atomic ensembles. The efficiency scaling allows to compare the different memory types but represents only one figure of merit.The applications in quantum information processing go beyond the simple analogy with classical memories where the signal is stored and retrieved. The different figures of merit should be considered in that prospect as the storage time, the bandwidth and the multimode capacity that we only superficially address when discussing the storage dynamics. In that sense, our contribution is mainly an introduction that can be pushed further to give a more complete comparison of the memories' performance.

Our objective was essentially to give a fundamental vision of few protocols that we consider as archetypes and hopefully stimulate the proposition of new architectures. We have finally replaced our analysis in the context of the quantum storage by deriving a variety of criteria adapted for both continuous and discrete variables. We have developed a toy model for the interaction of light with an atomic ensemble to evaluate the outcome of various quantum optics measurements that can serve as benchmarks to certify the quantum nature of optical memories.

We haven't insisted on the different material systems that physically represent the memory support. They all have in common to exhibit long lived (optical or spin) coherent states but they can cover different realities going from cold atomic vapors to luminescent impurities in solids as rare-earth doped insulators or excitons in semi-conductors holding a lot of promises in terms of integration. The portability of each protocol to a specific system would deserve a discussion by itself for which our analytic review of protocols can be seen as an introductory basis.

\section*{Acknowledgments}
Research at the University of Basel is supported by the Swiss National Science Foundation (SNSF) through the NCCR QSIT, the grant number PP00P2-150579 and the Army Research Laboratory Center for Distributed Quantum Information via the project SciNet.
The work at Laboratoire Aimé Cotton received funding from the national grant ANR DISCRYS (ANR-14-CE26-0037-02), from Investissements d'Avenir du LabEx PALM ExciMol and ATERSIIQ (ANR-10-LABX-0039-PALM).
The work at Laboratoire Pierre Agrain has been partially funded by the French National Research Agency (ANR) through the project SMEQUI.

\newpage
\appendix

\section{Strong pulse propagation}\label{strong_pulse}

Even if the standard 2PE presented in section \ref{2PE} is not appropriate for quantum storage, as an example, it illustrates a potential issue when strong pulses are used in echo sequences. The $\pi$-pulse as an element of the toolbox for quantum memories should be used with precaution. The best anticipated answer is certainly not to use the $\pi$-pulse as proposed in the controlled reversible inhomogeneous broadening protocol detailed section \ref{CRIB}. We here briefly discuss the propagation of strong $\pi$-pulses which appeared as critical element to understand the 2PE efficiencies as simulated in fig.\ref{fig:2PE_simul}.

We have assumed the $\pi$-pulse sufficiently short to have a well-defined action of the stored coherence. In practice, the rephasing pulse should be much shorter than the signal. This condition should be maintained all along the propagation which is far from guaranteed. $\pi$-pulses are very singular in that sense because they maximally invert the atoms irremediably associated to the lost energy from the pulse.  The requirement of energy conservation actually imposes a distortion of the pulse. There is no analytical solution to the propagation of strong pulses in absorbing media. Numerical simulations are then necessary to predict the exact pulse shape. That being said, the qualitatively analysis can be reinforced by invoking the McCall and Hahn Area Theorem \cite{area67, allen2012optical, Eberly:98}. This latter gives a remarkable conservation law for the pulse area though propagation as
\begin{equation}
\partial_z\theta(z)=
-\displaystyle\frac{\alpha}{2} \sin\left(\theta(z)\right)\label{area}
\end{equation}

In the weak signal limit (small area \cite{crisp1970psa}), one retrieves the Bouguer-Beer-Lambert law for the area (eq.\ref{bouguer}) as expected in the perturbative regime. A $2\pi$-pulse typically undergo the so-called self-induced transparency (SIT) \cite{area67}. The shape preserving propagation \cite{allen2012optical} is not surprising in the light of the energy and area conservations of $2\pi$-pulses. Indeed, a $2\pi$ Rabi flopping of the atoms doesn't leave any energy in the population. Additionally, the $2\pi$-area is unaffected by the propagation as given by the singularities of eq.\eqref{area} (for any area as multiple of $\pi$).

Along the same lines, a $\pi$-pulse conserves its area but not its energy. Pulse distortions are then expected to satisfy two contradictory conditions on the energy and the area. The pulse amplitude is reduced as the duration increases to preserve the total area. The energy scales as the pulse amplitude (multiplied by the area which is constant in that case) is then reduced. Theses considerations should not be underestimated when strong and more specifically $\pi$-pulses are used. To illustration the pulse distortion, we question the expression \eqref{etaPi} by performing a numerical simulation with different  $\pi$-pulse durations (see fig.\ref{fig:2PE_simul}). The deviation for the expected scaling precisely comes the $\pi$-pulse distortion as observed in fig.\ref{fig:2PE_simul} (top). This is an  intrinsic limitation of $\pi$-pulse when used in absorbing ensemble. This latter is fundamental and cannot be avoided by using a cavity to enhance the interaction with a weakly absorbing sample. Same distortions are expected in cavities \cite{gti}. The real alternative is the complex hyperbolic secant (CHS) pulse as discussed in section \ref{strong_pulse_rose}. This latter is not only robust under the experimental imperfections (as power fluctuation) but is also much less sensitive to propagation distortions \cite{Demeter}. Following our analysis, there is no constraint, as the area theorem, on the CHS as frequency swept pulses.


\section{Photon counting measurements}\label{appendix:formulas_counts}
We here give the detailed derivation of the formulas \eqref{autocorrelation}, \eqref{Cauchy_Schwarz} and \eqref{Bell} used in section \ref{counting_criterion}. We consider non photon-number resolving detectors with noise. Let $D_a(\eta_d)$ be the POVM element (positive-operator valued measure) associated to the event click when such a detector operates on a single mode of the electromagnetic field characterized by the annihilation $a$ and creation $a^\dag$ operators. Let $\eta_d$ be the efficiency of the detector and $p_{\text{dc}}$ the probability of a dark count. We have 
\begin{equation}
D_a(\eta_d) = \mathbb{1} - (1-p_{\text{dc}})(1-\eta_d)^{a^\dag a}.
\end{equation}
We first focus on the setup presented in fig.\ref{Fig1} by assuming that the noise and efficiency of the two detectors are the same. The ratio between the twofold coincidences and the product of singles is given by 
\begin{equation}
g_a^{(2)} = \frac{\langle D_{d_a}(\eta_d) D_{\bar{d}_a}(\eta_d)\rangle}{\langle D_{d_a}(\eta_d) \rangle \langle D_{\bar{d}_a}(\eta_d) \rangle}.
\end{equation}
Basic algebra using the relation between the modes $a,$ $d_a$ and $\bar{d}_a$ shows that $D_{d_a}(\eta_d) D_{\bar{d}_a}(\eta_d) = \mathbb{1}-2(1-p_{\text{dc}})(1-\eta_d/2)^{a^\dag a}+(1-p_{\text{dc}})^2(1-\eta_d)^{a^\dag a}$ and $D_{d_a}(\eta_d)=D_{a}(\eta_d/2)$. By including the memory efficiency in the detector efficiency, the ratio $g_a^{(2)} $ can be computed from
\begin{equation}
\label{auxg2}
g_a^{(2)} =\frac{\langle 1| \mathbb{1}-2(1-p_{\text{dc}})(1-\eta/2)^{a^\dag a}+(1-p_{\text{dc}})^2(1-\eta)^{a^\dag a} |1 \rangle}{\langle 1| \mathbb{1}-2(1-p_{\text{dc}})(1-\eta/2)^{a^\dag a}|1 \rangle^2}
\end{equation}
with $\eta=\eta_d\eta_m.$ Using an exponential form for $(1-\eta)^{a^\dag a}$ and expanding as a Taylor series, we find 
\begin{equation}
\label{meanvaluedet}
\langle n| (1-\eta)^{a^\dag a} |n\rangle  = (1-\eta)^{n}.
\end{equation}
Eq. \eqref{autocorrelation} is obtained by combining \eqref{auxg2} and \eqref{meanvaluedet}.\\

The expression for the Cauchy-Schwarz parameter is obtained from 
\begin{equation}
R=\frac{\langle D_{d_a}(\eta_d)D_{d_b}(\eta_d)\rangle^2}{\langle D_{d_a}(\eta_d)D_{\bar{d}_a}(\eta_d) \rangle \langle D_{d_b}(\eta_d)D_{\bar{d}_b}(\eta_d) \rangle}
\end{equation}
which leads to \eqref{Cauchy_Schwarz} by using the following results 
\begin{align}
\nonumber
\text{tr} (\rho_a x^{a^\dag a})& = \frac{1-p}{1-px},\\
\text{tr} (\rho_{ab} x^{a^\dag a + b^\dag b}) &= \frac{1-p}{1-px^2}
\end{align}
where $\rho_{ab}$ is the density matrix associated to a two-mode vacuum squeezed state and $\rho_a = \text{tr}_b \rho_{ab}.$

The expression for the visibility of the interference pattern observed in the Bell test experiment is obtained by noting that twofold coincidences are maximum between orthogonal polarizations while the minimum is obtained between identical polarizations. Hence, the numerator of eq.\eqref{Bell} can be obtained by taking the difference between $$\langle \psi^-_{a_h a_v b_h b_v} | D_{a_h}(\eta_d) D_{b_v}(\eta_d) | \psi^-_{a_h a_v b_h b_v}\rangle$$ and $$\langle \psi^-_{a_h a_v b_h b_v} | D_{ah}(\eta_d) D_{bh}(\eta_d) | \psi^-_{a_h a_v b_h b_v}\rangle$$ while the denominator comes from the sum of these two expectation values. \\

\newpage

\bibliography{AdvAMO_QMemories_bib}
\bibliographystyle{unsrt}

\end{document}